\documentclass{aastex6}
\usepackage{footmisc}
\usepackage{hyperref}
\usepackage{cleveref}

\shorttitle{Heating of an erupting prominence} 
\shortauthors{Lee et al.}

\begin{document}

\title{
Heating of an erupting prominence associated with a solar coronal mass ejection on 2012 January 27}

\author{Jin-Yi Lee$^{1}$, John C. Raymond$^2$, Katharine K. Reeves$^2$, Yong-Jae Moon$^{1,3}$, and Kap-Sung~Kim$^1$} 
\affil{$^1$Department of Astronomy and Space Science, Kyung Hee University, Yongin-si, Gyeonggi-do, 17104, Republic of Korea \\
$^2$Harvard-Smithsonian Center for Astrophysics, Cambridge, MA 02138, USA \\
$^3$School of Space Research, Kyung Hee University, Yongin-si, Gyeonggi-do, 17104, Republic of Korea \\
}  

\begin{abstract}
We investigate the heating of an erupting prominence and loops associated with a coronal mass ejection and X-class flare. The prominence is seen in absorption in EUV at the beginning of its  eruption. Later the prominence changes to emission, which indicates heating of the erupting plasma. We find the densities of the erupting prominence using the absorption properties of hydrogen and helium in different passbands. We estimate the temperatures and densities of the erupting prominence and loops seen as emission features using the differential emission measure method, which uses both EUV and X-ray observations from the Atmospheric Imaging Assembly on board {\it Solar Dynamics Observatory} and the X-ray Telescope on board {\it Hinode}. We consider synthetic spectra using both photospheric and coronal abundances in these calculations. We verify the methods for the estimation of temperatures and densities for the erupting plasmas. Then we estimate the thermal, kinetic, radiative loss, thermal conduction, and heating energies of the erupting prominence and loops. We find that the heating of the erupting prominence and loop occurs strongly at early times in the eruption. This event shows a writhing motion of the erupting prominence, which may indicate a hot flux rope heated by thermal energy release during magnetic reconnection.

\end{abstract}

\clearpage                                                                               

\section{Introduction}
Coronal mass ejections (CMEs) are among the most energetic solar events.
The CMEs are frequently associated with prominence eruptions \citep{munro1979, gosling1993, gilbert2000, gopalswamy2003, jing2004, schmieder2013, mccauley2015}. 
Statistical studies using white light coronagraph, EUV, and X-ray observations show that magnetic flux ropes, which exist before or are created during the CME eruptions in CME models (see references in \citealp{schmieder2015}), are a common occurrence in the evolution of CMEs \citep{vourlidas2013, nindos2015}.
The prominence eruptions and hot magnetic flux ropes associated with CMEs have been observed with high spatial and temporal resolution in extreme ultraviolet passbands since the {\it Solar Dynamics Observersvatory} (SDO) was launched in February 2010 \citep{patsourakos2013, tripathi2013, cheng2014, chen2014, amari2014, reeves2015, aparna2016, cheng2016a}. 

The energy budget is central to understanding the eruption process. Ideal MHD forces can eject plasma with little heating, while dissipation of magnetic free energy both heats and accelerates prominence material. Other processes such as dissipation of MHD waves, shock waves and ion-neutral interactions can also heat the gas. It is not trivial to determine the local heating rate even if the mass, temperature and speed of plasma are known, because thermal conduction can redistribute the energy and radiative or adiabatic cooling can keep the temperature low in spite of intense heating. However, it is possible to estimate the importance of their effects, and determine at least approximately how energy is partitioned among its kinetic, potential and thermal components.
                                                                                                                                                                       
EUV observations by Atmospheric Imaging Assembly (AIA) on board SDO show that the absorption features of prominences/filaments can change to emission features during their eruptions, which indicates that there is heating of the erupting plasma. It is well known that the CME plasmas experience strong heating in the earlier stage of their eruptions based on earlier studies using the observations by the Ultraviolet Coronagraph Spectrometer (UVCS) on board {\it Solar and Heliospheric Observatory} (SOHO) \citep{akmal2001, lee2009, murphy2011, lee2012} and the Advanced Composition Explorer \citep{gruesbeck2011, rakowski2011, lynch2011}.

Densities of prominences seen as absorption features in EUV have been estimated using the absorption properties of hydrogen and helium \citep{golub1999, gilbert2005, anzer2005, gilbert2006, gilbert2011, landi2013, williams2013, schwartz2015}. \citet{gilbert2005} find that the column density and mass of an eruptive prominence observed by the Extreme ultraviolet Imaging Telescope (EIT) are 1.6$\times$10$^{19}$ cm$^{-2}$ and 7.4$\times$10$^{14}$g, respectively. \citet{landi2013} find that the column densities of blobs of prominences in about 33000 K are about 3$\times$10$^{19}$ cm$^{-2}$ using observations by AIA. \citet{schwartz2015} have found column densities and mass of six quiet prominences using multi-spectral observations (EUV, X-rays, H$\alpha$, and CaII H) of about 10$^{18}$-10$^{19}$~cm$^{-2}$ and 2.9$\times$10$^{14}$ - 1.7$\times$10$^{15}$ g, respectively. Densities and temperatures of prominences seen as emission features in EUV have been estimated using differential emission measure methods (DEMs). 
\citet{kucera2008} investigate the heating of an erupting prominence using observations by the Solar Ultraviolet Measurements of Emitted Radiation (SUMER) on board SOHO and {\it Transition Region and Coronal Explorer} (TRACE).
Especially, \citet{gilbert2013} have investigated energy release from falling prominence material. They compare the kinetic energy of the prominence seen as absorption features with the radiated energy, and find that the dominant mechanism of energy release seen in emission features is plasma compression. 

Thermal structures of erupting plasmas have been investigated by differential emission measure (DEM)  methods. Multiple structural components show high temperature plasma ($>$ 8~MK) in the flux rope of a CME \citep{cheng2012}. A study of multi-thermal dynamics and energetics of a CME shows a hot erupting core in 11-14 MK \citep{hannah2013}. \citet{hanneman2014} and \citet{reeves2017} have investigated the thermal structures of current sheets and supra-arcade downflows using both AIA and XRT observations to measure the higher temperatures more precisely. 
Hot flux ropes ($>$ 10~MK) with a kinked configuration have been presented that are heated by a magnetic reconnection \citep{patsourakos2013, tripathi2013}. 

In this analysis, we investigate the heating and energetics of an erupting prominence and loop associated with an X class flare and a fast CME using the AIA and XRT observations. We estimate the densities of the erupting prominence using the absorption properties of hydrogen and helium at the beginning of the eruption. We estimate the temperatures and densities of erupting prominence and loop using a regularized DEM procedure \citep{hannah2012} modified to use both AIA and XRT observations. We then find the kinetic, thermal, radiative loss and thermal conduction energies. Lastly, we discuss the heating of the erupting prominence and loops.

In section 2, we describe the eruptions of prominence and loop used in this analysis. In Section 3, we explain the analysis methods to investigate the erupting plasmas as seen absorption and emission. In Section 4, we discuss the estimated physical properties (temperature, density, mass, heating, and energetics) of the erupting plasmas. In Section 5, we present our conclusions.

\section{Observations}
\label{sec:obs}

A fast halo coronal mass ejection was observed originating from the solar west limb on 2012 January 27. The CME is associated with an X1.7 flare observed 
by  {\it{Geostationary Operational Environmental Satellite}} (GOES) and has a linear speed of 2500km/s\footnote{\label{catalogue}http://cdaw.gsfc.nasa.gov/CME$\_$list/}.
Prior to the eruption, a prominence is observed and the flare starts at 17:37 UT, peaks at 18:36 UT, and ends at 18:56 UT\footnote{\label{flarelist}http://www.lmsal.com/solarsoft/latest$\_$events/}.
The AIA \citep{lemen2012} on board {\it SDO} \citep{pesnell2012} observes the solar corona and transition region up to 0.5~R$_{\sun}$ above the solar limb 
with high spatial ($\sim$0.6$''$ pixels) and temporal ($\sim$12 sec) resolutions in seven narrow EUV passbands (94 \AA, 131 \AA, 171 \AA, 193 \AA, 211 \AA,
304 \AA, 335 \AA). Full disk images are taken with a 4K$\times$4K CCD camera. 
AIA observed the prominence eruption on the solar west limb as absorption in all passbands of the AIA, except for the 304 \AA\ passband where it is observed as emission. 
During the eruption, the prominence shows a writhing motion and changes from absorption to emission features in all passbands. 
A loop which lies above the prominence appears in emission in 94 \AA\ and 131 \AA. 
This loop expands in 94 \AA\ and 131\AA\ but it is not seen in other lower temperature response passbands like 171 \AA, 193 \AA, 211 \AA, and 335 \AA.

The XRT \citep{golub2007, kano2008} on board {\it Hinode} \citep{kosugi2007} observes the solar corona with a spatial resolution of 2$''$ ($\sim$1$''$ pixels) and 
a field of view of 34$\times$34 arcmin. 
The XRT can automatically detect flares in the field of view of its 2K$\times$2K CCD.
In the event of a flare, the XRT observes the region of flare in a flare mode observing program.
For this event, the XRT takes images in flare mode between 17:47 UT and 17:58 UT with a 40 sec cadence and a field of view of 384$\times$384 pixels 
in Be\_thin, Al\_med, and Be\_thick passbands. XRT observes the overlying loop expansion in the Be\_thin and Al\_med passbands.
The observations in the Be\_thick passband have a signal to noise ratio that is too small to be useful. 
Thus, we use observations in the Be\_thin and Al\_med passbands. 
In the observations before the flare mode observation, the XRT observed the region where the event occurred with Ti\_poly and Be\_thin with a 150 sec cadence and a field of view of 384$\times$384 pixels. 

We remove the diffraction and scattering effects from the AIA observations. 
We use the standard point spread function\footnote{\label{flarelist}http://hesperia.gsfc.nasa.gov/ssw/sdo/aia/idl/psf/DOC/psfreport.pdf}, which includes the pattern of these effects, calculated using the procedure aia\_calc\_psf.pro at each passband of AIA.
Then, we deconvolve the Level 1.0 images with the point spread function using the procedure aia\_deconvolve\_richardsonlucy.pro to remove these effects. 
After the deconvolution, we calibrate the images to Level~1.5 by a standard procedure aia\_prep.pro that co-aligns and adjusts the plate scales and roll angles between AIA channels. 
For the XRT observations, we calibrate the images using a procedure (xrt\_prep.pro) in SolarSoft \citep{kobelski2014}. Then, we deconvolve the calibrated images using a point spread function supplied by the XRT team \citep{afshari2016}. 
We exclude the pixels that show contamination spots, which are spot-like patterns over the CCD that accumulated after the second bakeout on 3 September 2007 (see details \citealp{narukage2011}).

Figure 1 shows observations by AIA and XRT at four times. 
At 17:36~UT, the prominence is seen in absorption in the AIA observations.   
The prominence is more clearly seen in the lower temperature response passbands. 
It begins to rise very slowly from $\sim$17:10~UT and then starts brightening at $\sim$17:40~UT. 
However, near the bottom of the south leg, the prominence starts to brighten at $\sim$17:34~UT. 
We show the XRT observations using Ti\_poly and Be\_Thin at about the same time as the AIA observations. 
At this time, the XRT observations show the loops that surround the prominence and erupt at a later time. 
The loop over the prominence is also seen in the 94 \AA\ observation.

At 17:47~UT, the erupting prominence with a writhe structure is seen in emission in the lower temperature response passbands, 
while the erupting loop is seen in the higher temperature response passbands (94 \AA, 131 \AA, Be\_thin, Al\_med). 
The prominence is seen in emission, and begins to show writhing motions at $\sim$17:40~UT when it also begins to show brightening. 
The erupting loop seen in the higher temperature passbands is spatially co-located with the prominence and may consist of several loops along the line of sight. 

At 17:58~UT, the loop is erupting, and the prominence is no longer seen in the observations. The prominence appears to partially erupt or fall down until $\sim$17:55~UT.
At 18:08~UT, the erupting loop is outside the XRT field of view, and the XRT observations start again at $\sim$18:15~UT after the flare mode observations end at 17:58 UT. 
Thus, we show the erupting loop only in the AIA observations. The loops are more clearly seen in 131~\AA\ than 94~\AA\ for a longer time. 

\section{Analysis}

We estimate the densities of the erupting prominence seen in absorption using the absorption properties of hydrogen and helium. We estimate the temperatures and densities of the erupting prominence and loops as seen in emission using a DEM procedure \citep{hannah2012}. 
Then, we determine the masses of the erupting prominence and loops by assuming that their structures are cylinders. Finally, we find the kinetic, thermal, radiative loss, thermal conduction, and heating energies of the erupting prominence and the loops (Table~1). 

\subsection{Selection of regions seen as absorption and emission features}

 We select two regions to represent the cool prominence material.  
The 304 \AA\/ band may be contaminated by emission from the prominence or by
resonantly scattered $\lambda$ 304 \AA\ photons from the chromosphere \citep{labrosse2012}, so we do not use it in the absorption analysis.  
We select the region seen in absorption in all the other
AIA bands at the 17:36 UT.  Figure 2 shows the contours of the selected region.
For the second region we select the part of the prominence that shows writhing motion at 17:47 UT seen in Figure 3 (a).

For the erupting loops, we select regions at three times, 17:47~UT, 17:58~UT, and 18:08~UT, shown in Figure~3 (b)-(d).
The erupting loop at 17:47 UT includes the region selected for the writhing prominence viewed in emission. 
Therefore, the DEM of this region will represent the temperatures and densities of both the prominence and loop along the line of sight. 
We also show the selected regions for the erupting loops in the XRT/Be\_thin observations at 17:47~UT and 17:58~UT.
The selected region of the erupting loop at 18:08~UT is shown in 94 \AA. 

\subsection{Column density of a prominence seen as absorption features}

We determine the neutral and ion column densities of the erupting prominence materials as seen in absorption at 17:36 UT, shown in Figure~2.
First, we find the ratios (r$_{obs}$) of the observed emission (I$_{obs}$) absorbed by the prominence to 
 the emission observed without the existence of the prominence (I$_0$) for the region enclosed by contours in Figure~2.
 The images to estimate the I$_{obs}$ are shown in Figure~2 (a).
For the I$_{0}$, we use the averaged observations of 50 images for about 10 minutes between 17:10 UT and 17:19 UT for each passband.
During this time, the prominence eruption starts, but the prominence does not reach the full height of the prominence at 17:37 UT. The images to estimate the I$_0$ are shown in Figure~2 (b). 
 The ratios are calculated with the averaged values of the pixels within contours. These are shown with their uncertainties in Figure~4. 
Second, we find the neutral and ion column densities using an equation \citep{kucera1998, williams2013} given by 

\begin{equation}
I_{obs} = I_{b} (f_{clear} + f_{dark} e^{-\tau}) + I_{f} \\
= I_{0} - f_{dark} I_{b} (1 - e^{-\tau}), 
\end{equation}
\noindent where $I_0 = I_b + I_f$ and $f_{clear}=1-f_{dark}$. 
The $I_0$ includes the background emission ($I_b$) behind prominence and the foreground emission ($I_f$) between prominence and observer. 
We consider a covering factor, $f_{dark}$, and the fraction $f_{clear}$ accounts for partial covering of the emitting region behind the prominence if the prominence material is patchy.  
We assume a single value of the covering factor for both neutral and ionized particles. 
The equation can be written as

\begin{equation}
r_{obs}(\lambda_{AIA})=\frac{I_{obs}(\lambda_{AIA})} {I_{0}(\lambda_{AIA})}= 1 - G(1 - e^{-\tau(\lambda_{AIA})}), 
\end{equation}

\noindent where $G = f_{dark}I_{b}/I_{0}$. The G factor (0 $\leq$ G $\leq$ 1) combines the two unknown parameters, $f_{dark}$ and $I_{b}$. 
 The $\lambda_{AIA}$ are the AIA passbands (94~\AA, 131~\AA, 171~\AA, 193~\AA, 211~\AA, 335~\AA). 

The opacity $\tau(\lambda_{AIA})$ is defined as 

\begin{equation}
\tau(\lambda_{AIA}) = N_{neu}(\sigma_{H I}(\lambda_{AIA}) + A_{He} \sigma_{He I}(\lambda_{AIA})) + N_{ion} A_{He} \sigma_{He II} (\lambda_{AIA}),
\end{equation}

\noindent where N$_{neu}$ and N$_{ion}$ are the neutral and ion column densities, respectively. The A$_{He}$ term is the helium abundance from (N$_{He}$/N$_H$) and we use 0.085 \citep{grevesse2007}. 
Estimates of helium abundances in prominences range from 0.05 \citep{iakovkin1982} to 0.2 \citep{delzanna2004} and probably vary within each prominence \citep{gilbert2007}. 
The $\sigma_{H I}$, $\sigma_{He I}$, and $\sigma_{He II}$ are the absorption cross sections of neutral hydrogen, neutral helium, and ionized helium, respectively \citep{reilman1979, burgess1965}. 
The ionization energy of He~II is 54.4 eV which corresponds to 228 \AA. 
Therefore, the He II absorption does not affect the 304~\AA\ and 335~\AA\ passbands, so that it is possible to constrain the ion column density. 
These absorption properties have been used to estimate the temperatures and densities of prominences \citep{gilbert2011, landi2013}.

Using the above equations, we compute the ratio with the grids of the neutral and ion column densities in the range of 
10$^{17} - $10$^{22}$~cm$^{-2}$ for each of the AIA bands, except for 335 \AA\ because of the large range of wavelengths it covers. 
The 335 \AA\ attenuation depends on the DEM of the background emission. 
We assume the attenuation factor for 335~\AA\ is given by 

\begin{equation}
e^{-\tau(\lambda_{335})} = \frac{\sum_{ \lambda=125}^{800}{A_{eff}~{S(\lambda)}~e^{(-\tau(\lambda))}} } {\sum_{ \lambda=125}^{800}{A_{eff}~{S(\lambda)} } }
\end{equation}

\noindent where S($\lambda$) is the spectrum computed by CHIANTI with its Active Region DEM and A$_{eff}$ is the effective area for 335~\AA.
 
We assume that the 1 $\sigma$ error = (the difference between lower and upper bounds to ratio for each band)/2.
The lower and upper bounds are computed using the DN errors by a standard procedure (aia\_bp\_estimate\_error.pro in SSW) with the evenorm keyword only, 
which includes the systematic error due to normalizing the effective area using cross-calibration with the EUV Variability Experiment (EVE) on board SDO. 
This error estimation gives the uncertainty in the measurement (DN/pixel). 
However, this method uses the relative intensities which are not affected by the systematic errors due to the normalizing the effective area and uncertainties in the preflight photometric calibration and CHIANTI data. 

We compute $\chi^2$ given by

\begin{equation}
\chi^2 = \sum{(r_{comp}(\lambda_{AIA}) - r_{obs(\lambda_{AIA})})/\sigma_{error}(\lambda_{AIA})^2},
\end{equation}

\noindent over a 3-dimensional grid consisting of factor (G), neutral column density (N$_H$), and ion column densities (N$_{ion}$), respectively.
We find the region which satisfies the condition, $\chi^2$ - minimum($\chi^2$)~$<$ 3~$\sigma_{error}$. 
Here, the minimum of $\chi^2$ represents the uncertainties that could affect the observations due to the changes of background emissions for different passbands 
as well as the large range of wavelength of 335 \AA\ observations. 

A study of a moving prominence jet in a EUV using a DEM analysis showed that the jet is a multi-thermal
structure with plasma temperatures at least as low as LogT=5.0 and probably as high as LogT=5.8 \citep{kucera2006}. 
\citet{parenti2012} show that the prominence-corona transition region has significant plasma emitting at $>$ 4$\times$10$^5$ K using AIA observations. 
There is an open issue whether the high temperature part of prominence DEMs is the result of the foreground and background emission \citep{parenti2014}.  
A basic limitation of the method used here is that there may be a component of emission associated with the prominence material. 
Indeed, at later times, the prominence is seen in emission instead of absorption. 
Even at earlier times, these are seen in emission in the coolest temperature band, He II 304 \AA.  
We do not have enough observables to rule out emission from the prominence in the other AIA bands, 
but note that unless that emission happened to mimic the radiative intensities of the I$_0$ in the various bands, it would show up as scatter in values of $\tau$ 
derived for the different bands. The strongest constraint would be for the AIA 171\AA\ and 193 \AA\ bands, for which the opacities are nearly the same. 
If there is a contribution of prominence emission, it would lead to an underestimate of $\tau$ and the column density, 
but based on the agreement between $\tau_{171}$ and $\tau_{193}$, such a contribution is probably small. 

\subsection{Temperatures and densities of the erupting prominence and loops seen in emission}

We estimate the temperatures and densities of the erupting prominence and loops seen in emission using a code to produce DEM maps, which performs a zeroth order regularization. 
This code is an optimized and faster version of a DEM code\footnote{\label{dem}http://www.astro.gla.ac.uk/$\sim$iain/demreg/map/} developed by \citet{hannah2012, hannah2013}. 
The main code (dn2dem\_map\_pos.pro) uses the observations of six passbands (94~\AA, 131~\AA, 171~\AA, 193~\AA, 211~\AA, 335~\AA) in AIA as input and returns the DEM maps and errors.
The code calculates the uncertainties of observations in the code itself for the DEM calculations.
The uncertainties include read noise for each passband and shot noise in photon counting (the square root of the number of photons) \citep{boerner2012}.
In this analysis, we use only the uncertainties in read and shot noise, because other uncertainties such as those in the preflight photometric calibration and CHIANTI data are poorly known. 
As discussed below, these systematic uncertainties are significant for 94~\AA\ and 335~\AA\ bands. 

We revise the zeroth order regularization code (dn2dem\_map\_pos.pro) to use with the XRT observations. 
Then, we use the observations of six passbands in AIA and two passbands (Be$\_$thin and Al$\_$med) in XRT for the DEM.
We use the AIA observations with the same field of view as the XRT observations of about 394$''\times$394$''$. 
The plate scale of  the AIA observations (0.6 $''$ pixel) is rescaled to match the XRT plate scale (1.0286$''$ pixel). 
Also, we extend the temperature range to 0.1~MK$-$100~MK from 0.5 MK$-$32 MK which used in the original code. 
For the uncertainties of the XRT observations, we use a formula given by 

\begin{equation}
\rm{\xi (XRT)} = (1.0 + \sqrt{(\rm{observations(DN) + 0.75})} ) / \rm{exposure~time}, 
\end{equation}
 
\noindent which tends to Gaussian for high counts and Poissonian for low count regimes \citep{gehrels1986}. 
We take pixels above $1.5\times 10^4$ to be saturated and pixels with Signal/Noise $<$ 3 to be noise-dominated.  
Those pixels are not used in the DEM calculation.  
For XRT we take the saturation level to be 2500 DN.

DEM calculations use temperature response functions that can be made with different elemental abundances.
Elements with low first ionization potential (FIP $<$ 10~eV) are enhanced in the corona by about a few factors with respect to the photosphere \citep{fludra1999, grevesse2007}. 
In a review of prominences, \citet{labrosse2010} have discussed abundances of solar prominence. 
Studies of prominences and emerging flux from the observations by {\it Skylab} show that the abundances are between photospheric and coronal \citep{widing1986, spicer1998, sheeley1995}. 
The eruption in this analysis is associated with a prominence. 
Thus, the abundances of the elements in the prominence are probably closer to the photospheric abundances than the coronal abundances.  
We investigate whether the reconstructions from the DEM calculation can be improved by using different elemental abundances. 
We use four kinds of spectra using different elemental abundances and versions of CHIANTI, since Version 8 incorporates new atomic rates \citep{delzanna2015}. 

Firstly, we use the default AIA spectra response (aia\_get\_response Version 6 in SSW)\citep{boerner2014}, which uses a coronal abundance 
(sun\_coronal\_1992\_feldman\_ext.abund in SSW, hereafter Feldman abundance)\citep{feldman1992, landi2002, grevesse1998} from CHIANTI~7.1.3. 
We also make a temperature response function using the standard procedure (aia\_get\_response.pro in SSW) with several options, chiantifix, evenorm, and noblend.
The other three spectra are made using CHIANTI~8 \citep{delzanna2015} applying three different abundance sets. 
For the second spectra, we use the Feldman abundance, which is the same as the default spectra of the AIA, but the spectra are made using CHIANTI~8. 
For the third spectra, we use a different coronal abundance (sun\_coronal\_2012\_schmelz.abund in SSW)\citep{schmelz2012}.  
This abundance takes into account that low-FIP and high-FIP elements are enhanced by a factor of 2.14 and 0.71, respectively in the corona relative to the photosphere 
while the Feldman abundance takes that low FIP elements are enhanced by a factor of four with respect to their photospheric values and high-FIP elements are the same in the corona and the photosphere. 
For the fourth spectra, we use a photospheric abundance (sun\_photospheric\_2011\_caffau.abund in SSW)\citep{caffau2011, lodders2009} since the prominence may contain plasma with photospheric abundances. 
These abundance sets probably do not cover the full range of possibilities. For example, \citet{mohan2000} found large variations in the Mg/Ne ratio within a single prominence. 
We show the temperature response curves in Figure~5. The AIA responses are, not surprisingly, simply proportional to the Fe abundance since the emissions observed by the AIA mostly come from Fe ions. 
The XRT response has a strong Fe contribution, but it also includes contributions from the low-FIP elements Mg and Si, the high-FIP element O, and from bremsstrahlung. 

\subsection{Geometry and Mass estimation of the erupting prominence and loops}

We estimate the densities and masses of the erupting prominence and loops by assuming their structures to be cylinders. 
The width and length of the prominence seen as absorption features are shown at 193 \AA\ in Figure~2~(a). 
The widths and lengths of the erupting prominence and loops seen as emission features are shown in Figure~3. 
We assume that the width of the cylinder structure is equal to the line of sight depth. 
The depths and lengths are shown in Table~2. 
In the case of the erupting prominence in Figure~3~(a), the length is multiplied by 2 because it shows only the half of the loop length.

For the prominence as seen in absorption, we estimate the total column density by the summation of the neutral and ion column densities for each pixel in the grid satisfying the constraint in Section~3.2. 
The covering factor, $f_{dark}$, can be taken to be a lower limit ($f_{dark} > G$) since the term $I_b/I_0$ is less than 1 and the factor $G$ is between 0 and 1 (Kucera et al. 1998). Therefore, we estimate the column density of the prominence multiplying by $f_{dark}$ as a lower limit. 
Based on Figure~4, $I_b/I_0$ must be greater than 0.4. Therefore,  the column density of the prominence could be up to 2.5 times our estimate.  
Then, we estimate the density of the prominence by dividing the column density by the line of sight depth. 
Assuming a cylinder structure of the prominence, we estimate the mass as a lower limit. 

For the erupting loops seen as emission features, we estimate the density using the DEM explained in Section~3.3 by a relationship below.

\begin{equation}
EM = < n_e^2 > dl = \int {DEM} dT,
\end{equation}

\noindent where n$_e$, EM, and dl are the electron number density, emission measure, and line of sight depth, respectively. 
We assume the filling factor to be 1 for the erupting loops. 
This provides the upper limit to the masses of the observed eruptive plasmas since
any clumping will cause an overestimate of the mass. 
We assume a plasma with a 10$\%$ of Helium content with a mass of 1.97 $\times$10$^{24}$ g per particle.

\subsection{Energetics and heating of the erupting prominence and loops}
We estimate the thermal, kinetic, radiative loss, thermal conduction, and heating energies and the radiative loss and thermal conduction time scales (Table~1, see details Section 3.3 in \citealp{lee2015}).
Also, we estimate the energy loss rates by radiation, and thermal conduction and heating rates of the erupting loops. 

We estimate the heating energy by the energy equation given by

\begin{equation}
\frac{5}{2} (n_e+0.8n_e)k\frac{dT}{dt} = -(dq + L_{rad} +  L_{a} - H_r), 
\end{equation}

\noindent where dq, L$_{rad}$, and L$_{a}$ are cooling rates by thermal conduction, radiation, and adiabatic expansion, respectively. 
The heating rate, H$_r$, includes all contributions to the heating, such as wave dissipation, ohmic heating, turbulent heating, the divergence of conductive flux, or shock waves generated by the reconnection outflow. The heating energy (H) is estimated by multiplying H$_r$ by the volume of the erupting prominence and loop and $dt$. 
We assume that the proton density is 0.8 times n$_e$.
The k and T are Boltzmann constant and temperature, respectively. 
L$_{rad}$ is estimated by 

\begin{equation}
L_{rad} = 0.8 n_e n_e P(T), 
\end{equation}

 \noindent where P(T) is plasma radiative loss function \citep{klimchuk2008}.  
 Assuming the specific heat ratio $\gamma$=5/3, the L$_{a}$ can be expressed as  
 
 \begin{equation}
L_{a} = \frac{5}{2}(n_e + 0.8 n_e) k (2T_0\frac{v_h}{h}), 
 \end{equation}

\noindent where T$_0$ is the temperature of the prominence and erupting loops in the previous time, v$_h$ is the expansion velocity, and h is the height of the loop. 
We define the height from the solar limb to the loop top in the observations as represented in Figure~3.  
For v$_h$, we estimate the expansion velocities of the prominence and loops at 193~\AA\ and 131~\AA\ observations, respectively. 

We assume T$_0$ for the prominence as 80000 K since it is seen as emission at 304~\AA\ (see Section 4.1). 
For the loop at 17:47~UT, we estimate the temperature of the loop by the DEM analysis at 17:36~UT using the observations in the first columns in Figure~1. 
At this time, we use Ti\_poly and Be\_thin observations because Be\_thin and Al\_med observations start at 17:47~UT. 
We find a temperature around 6~MK and use it for the T$_0$ for this loop. 
For the loops at 17:58~UT and 18:08~UT, we use the temperatures of the loops estimated using the DEM analysis in the previous time at 17:47~UT and 17:58~UT, respectively, 
which will be described in Section~4.3. The $T_0$ is also used to estimate the temperature differences in the energy equation for $dT$(=T-T$_0$). 
The $dt$ is the time difference between the times used in this analysis. 
It is also used for the duration time ($\triangle{t}$) to estimate radiative loss and thermal conduction energies (see \citealp{lee2015}). 
The cooling rate from the thermal conduction, $dq$, is estimated using thermal conduction energy divided by the volume of the erupting prominence and loop and $dt$.
These parameters for the estimation of the energies are represented in Table~2.  

\section{Results and Discussion}
\label{sec:results}

We investigate the temperatures and densities of the erupting prominence and loops as seen in absorption and emission. 
Using these physical properties, we estimate and discuss the masses, energetics, and heating of the erupting prominence and loops. 

\subsection{Column density of the erupting prominence as seen in absorption}

Figure~6 shows the allowed range of column densities for the absorption features with the covering factor of 0.33$-$0.48. 
The ion column density is hard to constrain because 304 \AA\ and 335 \AA\ are not absorbed by He II, but we don't use the observations in 304~\AA, and 335~\AA\ has a significant response at short wavelengths and a relatively poor signal-to-noise. 
We find that the total column density of the erupting prominence is in the range of 4.0$\times$10$^{18}$~cm$^{-2}\sim~$9.2$\times$10$^{19}$~cm$^{-2}$ as a lower limit. 
 The column densities estimated in the previous studies \citep{gilbert2005, landi2013, williams2013, schwartz2015} are similar to our values. 
Assuming the line of sight depth of the prominence in Table~2, the number density is in the range between 7.6$\times$10$^9$ and 1.8$\times$10$^{11}$ cm$^{-3}$.

\citet{williams2013} find the column density of $\sim$10$^{20}$~cm$^{-2}$ for a fast moving prominence by a polychromatic method using the AIA observations in five EUV passbands except for 304~\AA\ and 335~\AA, 
and 10$^{18}$-10$^{19}$~cm$^{-2}$ as a lower limit by a monochromatic method using only one passband of 193~\AA. 
In their analysis, the comparison of two methods shows that the monochromatic estimates are systematically lower than those derived from the polychromatic method. 
\citet{landi2013} introduce a new diagnostic technique to determine the electron temperature and column density using AIA observations during 
prominence eruptions using the fact that He II can absorb only below 228 \AA , 
while H I and He I absorb all the EUV bands of AIA, along with the 
equilibium fractions of He vs. temperature.  It is sensitive in the range 
20,000 to 80,000 K where He I is ionized to He II.  Figure 6 shows that the 
AIA data are consistent with $N_{ion} / N_{neu}$ from essentially zero to 
about 35, which suggests $T \leq 70,000$ K. The erupting prominence 
analyzed here shows emission at 304 \AA , suggesting T $\sim$ 80000 K if it 
arises from collisional excitation.  For comparison, \citet{labrosse2012} found central temperatures of 6000 to 12,000 K in erupting 
prominences, assuming that the 304 \AA\/ emission is mainly chromospheric 
light resonantly scattered by the cool prominence material. However collisional 
excitation in the hotter surface layers likely dominates in the events 
where the 304 \AA\/ emission brightens as the velocity increases.

\subsection{Comparison of the observations and the reconstructions from DEM}

To assess the reliability of the DEM, we compare the reconstructed DN from the DEM analysis with the observations of the erupting prominence and loops for the regions shown in Figure~3. 
The reconstruction errors for model spectra calculated with various abundance tables and atomic data are shown in Figure~7. 
The comparisons of the reconstructions and the observations show that the reconstructions for 94 \AA\ and 335 \AA\ fit the observations poorly.  
Previous studies also show that the reconstruction errors at 94 \AA\ and 335 \AA\ are relatively larger than at other passbands \citep{hanneman2014, gou2015}.
The DEM errors using the spectra (triangle) predicted by CHIANTI~8 and the Feldman abundance are smaller than those using the spectra (diamond) predicted by CHIANTI~7 and the same coronal abundance for the 94 \AA\/ band, apparently because of the improved atomic rates. 
We find that the CHIANTI 8 improves the fit, particularly for the 94 \AA\ channel, but it does not entirely fix the problem.  
We cannot tell whether the remaining residuals result from calibration problems, atomic data problems or some physical effect such as non-Maxwellian electron distributions that is not included in the models.

The DEM errors for the XRT passbands with photospheric abundance by \citet{caffau2011} and CHIANTI~8 are a little smaller than with the AIA default spectra, which uses the coronal abundance by \citet{feldman1992} and CHIANTI~7. 
A possible reason is a difference between the actual abundances and those assumed, because the AIA count rates are proportional to the iron abundance, 
while the XRT count rates include contributions from Bremsstrahlung and lines of other elements. 
The AIA responses are simply proportional to the Fe abundance as shown in Figure~5. So the derived EM goes as 1/[Fe] and the derived density as 1/[Fe]$^{1/2}$.
The XRT bands are dominated by emission from the low-FIP elements Mg, Si and Fe, especially at high FIP bias.  Therefore, the choice of abundances does little to solve the discrepancy between AIA and XRT. 

\subsection{Temperatures and densities of the erupting prominence and loops as seen in emission}

We estimate the temperatures and densities of the erupting prominence and loops as seen in emission using the DEM method \citep{hannah2012, hannah2013} at three times.
Figure~8 shows the DEM maps of the erupting prominence and loops at these times. 
We show the pixels that satisfy the DEM error $\leq$ 30$\%$ and the temperature error $\triangle$LogT $\leq$ 0.25, 
which can be considered the uncertainties in the DEM analysis \citep{hannah2013}. 
We estimate the densities of the erupting prominence and loops with those pixels.
However, Figure~9 shows the DEMs for all pixels. The largest errors are the temperature errors in high temperature regions. 

In Figure~8 (a), the writhe structure of the erupting prominence is seen in the maps at 0.5$-$3 MK while the erupting loop is seen at higher temperatures $>$ 6~MK at 17:47~UT. 
The prominence is observed at 171 \AA, 193 \AA, 211 \AA, and 335 \AA, but not at 94 \AA\ and 131 \AA, while the loop is observed at 94 \AA, 131 \AA, Be$\_$thin, and Al$\_$med, but not at the other passbands (see Figure~1).
The DEMs of the prominence and loop for the regions represented in Figure~3 at 17:47~UT are shown in Figure~9 a) and b), respectively. 
The DEMs of the two regions are similar to each other because the regions selected overlap. 
The region for the erupting loop contains the prominence at this time. 
The DEMs peak at about LogT=6.3-6.4, which is similar to the coronal temperature, and also at around LogT=7.0. 
However, the DEM maps in Figure~8 (a) show clearly that the peak at about LogT=6.3-6.4 represents the erupting prominence with the writhe motion, 
and the peak at LogT=7.0 represents the DEMs of the loop. 
Thus, we estimate the densities of the prominence and loop by the summation of the EMs, which is multiplied by the temperature bins,  in 0.5$\sim$3~MK and 6$\sim$14~MK, respectively. 

The emissivities in the AIA bands are enhanced by about a factor of 4 with coronal abundances (enhanced in low-FIP elements such as Fe) 
compared to photospheric abundances \citep{feldman1992}.
As we discussed in Section 3.3, the density with photospheric abundance is enhanced about a factor of two relative to the density with the Feldman abundance. 
Assuming the line of sight depth of the prominence and loop in Table~2 and using various spectra, the estimated densities for the erupting prominence and loops are in the range of 1.3$\times$10$^{10}-$2.5$\times$10$^{10}$cm$^{-3}$ and 9.0$\times$10$^{9}-$1.6$\times$10$^{10}$cm$^{-3}$, respectively. 
The prominence absorption provides only loose constraints. 
However, the density of the prominence in emission is smaller than the density as seen in absorption. 
It might be due to the expansion of the prominence during the eruption and also the presence of the lower temperature plasmas which are not heated to be seen in emission. 
                                   
At 17:58~UT, the prominence is not seen anymore in the lower temperature maps while the erupting loop is seen in the higher temperature maps in Figure~8~(b).
The DEMs show the peaks at around LogT=6.3-6.4 and LogT=6.8-6.9 in Figure~9~(c). 
We determine the density of the erupting loop using the summation of the EMs in 4$\sim$11~MK because the peak at lower temperature might be the coronal background. 
We define the temperature of the loop as 8~MK. 
This is a little lower than the temperature at the previous time. 
The estimated density is 8.6$\times$10$^{9}-$ 1.6$\times$10$^{10}$cm$^{-3}$. 
The density is similar to the density at the previous time.

At 18:08~UT, we find the temperature and density with DEMs using the AIA observations only. 
The high temperature component has faded dramatically. 
The DEM maps show that the erupting loop is seen in the broad temperature bands in Figure~8~(c). 
In the case of this loop, we determine the density by the summation of the EMs in 1$\sim$14~MK. 
The estimated density is 6.5$\times$10$^{9}-$ 1.2$\times$10$^{10}$cm$^{-3}$. 
The density of this loop is smaller than the previous time, but the mass is bigger due to the larger geometry from expansion. 
We use a temperature of 4~MK for this loop for the energy estimation in the next section. However, the loop spans a large temperature range. 
We show the temperatures, densities, and masses of the erupting prominence and loops  in Table~3.

The derived densities of the erupting loops are larger than the densities $\sim$1$\times$10$^{9}$cm$^{-3}$ of the core regions of CMEs in previous studies \citep{cheng2012, hannah2013}. 
The event analyzed in this paper is associated with a large CME event 
which has a mass 3.7$\times$10$^{16}$g\footref{catalogue}, while the masses of the events in the previous studies are about 1$\times$10$^{15}-$6$\times$10$^{15}$g\footref{catalogue}. 
Therefore, it is possile that the density of the erupting loops in the lower corona associated with the massive CME is higher, 
although we can not exactly say whether the CME is denser or just larger than these other events. 

The DEMs show a bump at higher temperatures ($> \sim$LogT=7.5). However, the temperature errors are larger. 
One possible reason might be due to the continuum observed by XRT.
The analysis using various spectra does not improve this problem. 
In this analysis, we use the zeroth order regularization procedure \citep{hannah2012} and use only two passbands for the XRT observations. 
A comparison of the DEMs of a flare using the AIA only and AIA + XRT, which use the three passbands of the XRT observations, shows that the DEMs constructed by adding the XRT observations show a narrower peak at hot temperatures \citep{hanneman2014}. 
It might be worth investigating the DEMs using the larger number of the XRT passband observations in the future. 

\subsection{Energies and heating}

Table~3 shows the energetics of the erupting prominence and loops.
The eruptions of the prominence and loops are associated with an X-class flare and a powerful fast halo CME which has the kinetic energy of 1.2$\times$10$^{33}$ erg\footref{catalogue}. 
The kinetic energies of the prominence and loops are much smaller than a kinetic energy of the CME. 
The speeds and masses of the erupting loops are smaller than those of the CME at 33$-$85~km/sec and about 1$\times$10$^{15}-$2$\times$10$^{15}$g, respectively. 
The erupting plasma might not be accelerated at this low height of about $\sim$10$^5$~km, and 
the masses in EUV and X-rays are smaller than the total CME mass observed by the Large Angle and Spectrometric Coronagraph Experiment (LASCO) on board $\it{SOHO}$, 
which includes cooler material.  

For the prominence at 17:47~UT, the thermal energy (TE) and heating energy (H) are similar to each other. 
With the relatively lower temperature and higher density of the prominence, the radiative loss energy (L$_r$) is larger than thermal conduction energy (F$_c$). 
Also, the cooling caused by radiative losses is larger than the cooling by adiabatic expansion and thermal conduction. 
The heating rate (H$_r$) is a few times larger than the sum of energy loss rates. 
The radiative loss rate is relatively large due to the high density of the prominence. Thus, the observed emission requires the high heating rate. 
If the heating rate continues to be larger than the sum of the cooling rates, then the prominence material will probably reach even higher temperatures. 
In the case of the loop at 17:47~UT, the thermal energy is a little larger than the heating energy. 
With the high temperature of the loop, the thermal conduction energy is larger than the radiative loss energy, and the thermal conduction timescale is shorter than the radiative loss timescale. 
On the other hand, the cooling by the adiabatic expansion is larger than the other cooling terms. 
The heating rate of this loop is also larger than the sum of the cooling rates.  

For the loop at 17:58~UT, 
the energetics of the loop are still dominated by the pre-eruption energy since the thermal energy is larger than the heating energy. 
The radiative cooling and thermal conduction time scales are comparable to each other. 
In addition, the cooling rate of the adiabatic expansion is also comparable to the heating rate. 
The loop at 18:08~UT shows a multi-thermal structure as shown in the DEM maps in Figure~8. We estimate the energies assuming a loop temperature of 4~MK. 
The thermal energy is larger than the heating energy, and the cooling rate by adiabatic expansion is larger than the heating rate. 
It is hard to say how much the loop plasma in the previous time is being heated or cooled. 
Table~4 shows the energies using the densities of the loops in 6-14~MK at 17:58~UT and 18:08~UT. 
This temperature range is the same one used to estimate the density of the loop at 17:47~UT. 
We estimate the energies assuming also the same temperature of 10~MK at 17:47~UT. 
The density of the loop at 18:08~UT is about a half of the density of the loop at 17:58~UT in this hot temperature range. 
The thermal energies are larger than the heating energies at all three times. 
The cooling term by adiabatic expansion is the largest compared to the other cooling terms. 
The heating rates and energies become smaller with the loop expansion. 

Overall, at the earliest time of the eruption at 17:47~UT, the heating rates of both prominence and loop are larger than the cooling rates. 
In contrast, at the later time at 17:58~UT, the adiabatic cooling rates are comparable to the heating rate. 
Hot flux ropes have been observed with a kinked configuration \citep{patsourakos2013, tripathi2013}. 
\citet{tripathi2013} present an erupting prominence that shows a kinked flux rope that is heated by thermal energy release from magnetic reconnection. 
\citet{patsourakos2013} present an arcade magnetic field that forms a flux rope via successive reconnections. 
In this analysis, the heating of the erupting prominence and loop occurs strongly early in the eruption. 
Then, the erupting loops partially cool down due mostly to adiabatic cooling, but still remain fairly hot at later times. 

We compare the heating rates with the results in a previous study using the UVCS observations at 2.4~R$_\sun$ \citep{lee2009}. 
They find the heating rates with the radiative and adiabatic cooling rates using a time-dependent ionization state model 
assuming that all CME materials erupt at the same time and location and reach the UVCS slit with a constant speed. 
The heating rates in this analysis are larger than the results shown in their paper, which are computed with the initial temperature of 1.6$\times$10$^6$~K and density of 5$\times$10$^8$~cm$^{-3}$ at the beginning of the eruption in their simulation. In this analysis, the densities of erupting plasmas are higher, and we assume ionization equilibrium. 
Thus, it might not be appropriate to compare this eruption to previous results.  
Nevertheless, we find similar accumulated heating energies to those in \citet{lee2009}. They find the accumulated heating energies $\sim$10$^{15}$ erg/g during $\sim$700 sec at the beginning of the eruption. 
In this analysis, we find that the heating energies of the prominence and loop at 17:47~UT are $\sim1\times10^{15}$ erg/g and $\sim3\times10^{15}$ erg/g dividing by their masses during 648 sec. 
Similarly, \citet{landi2010} investigated the heating energy of erupting CME plasma using the EUV Imaging Spectrometer (EIS) and XRT on board Hinode and UVCS observations. 
They find that the required heating energies are $3\times10^{15}$ erg/g at 1.1 $R_\sun$ and $\sim7\times10^{14}$ erg/g at 1.9 $R_\sun$. 
Thus the heating energies are similar in the three events. 

At 17:47 UT, there was a minor peak in X-ray flux from GOES, 
but it was 50 minutes before the peak of the X-class flare. 
The X-ray energy loss rate was $\sim5\times10^{25}$ erg/s in the 1\AA\ $-$ 8\AA\ band, 
while the heating rates of the prominence and loop are $\sim5\times10^{26}$ erg/s and $\sim7\times10^{27}$ erg/s, respectively, in this analysis. 
Thus relatively little plasma was hot enough to produce X-rays above 1.5 KeV. 
Shortly afterwards the X-ray energy loss rate increased by a factor of 30, and 
the kinetic energy of the CME increased as the CME mass increased and the CME accelerated. 

\section{Summary and Conclusion}
\label{sec:conclusion} 

We investigate the heating of erupting prominence and loops associated with a coronal mass ejection on 2012 January 27. 
This event shows the eruption starting with the prominence eruption as seen in absorption, and then the erupting prominence/loops as seen in emission. 
These observations make it possible to investigate the heating of CME plasmas at early times during its eruption. 
We  constrain the neutral and ion column densities of the erupting prominence as seen in absorption by hydrogen and helium. 
We find that the total column density of the prominence is in the range of 4.0$\times$10$^{18}$~cm$^{-2}\sim~$9.2$\times$10$^{19}$~cm$^{-2}$ as a lower limit. 
We estimate the temperatures and densities of the erupting prominence and loops as seen in emission using a DEM method with various spectra. 
The comparisons of the reconstructions and the observations show that the reconstructions for 94~\AA\ and 335~\AA\ bands fit the observations poorly, 
although Version 8 of CHIANTI improves the fit for the 94~\AA\ band. The choice of abundances does little to solve the discrepancy between the AIA and XRT. 

We find that the temperatures of the erupting prominence and loop at the earlier time are about 2.5~MK and 10~MK, respectively. 
The thermal energies are comparable to the heating energies at the earlier time, and the heating rates are larger than any other cooling terms. 
This event shows a writhing motion of the erupting prominence, which may indicate a hot flux rope heated by energy release during magnetic reconnection. 
Also, the heated plasmas might partially cool down, but still remain fairly hot at later times. 
The accumulated heating per gram is comparable to estimates derived for other CMEs. 
The heating rates are larger than the results in \citet{lee2009}.  
This may indicate that the heating in this analysis is more localized in the earliest time with a low expansion speed. 
The observations pertain to a very early stage in an X-class flare. 
In the future, it will be worth investigating the heating of the erupting plasmas using other events such as this. 

\acknowledgments

This work was supported by Basic Science Research Program
(NRF-2013R1A1A2058409 and NRF-2016R1A6A3A119\\32534) and 
the BK21 plus program through the National Research Foundation (NRF) funded by the Ministry of Education of Korea, 
NRF of Korea Grant funded by the Korean Government (NRF-2013M1A3A3A02042232 and NRF-2016R1A2B4013131), 
the National Radio Research Agency/Korean SpaceWeather Center,  
and the Korea Astronomy and Space Science Institute under the R\&D program 
`Development of a Solar Coronagraph on International Space Station 
(Project No. 2017-1-851-00)' supervised by the Ministry of Science, ICT and Future Planning. 
KR is supported by NSF SHINE grant AGS-1156076 to the Smithsonian Astrophysical Observatory. 
``Hinode is a Japanese mission developed and launched by ISAS/JAXA, with NAOJ as
domestic partner and NASA and STFC (UK) as international partners. It is
operated by these agencies in co-operation with ESA and the NSC (Norway)." 
The CME catalog is generated and maintained at the CDAW Data Center by 
NASA and The Catholic University of America in cooperation with the Naval Research Laboratory. 
SOHO is a project of international cooperation between ESA and NASA. 

\bibliographystyle{aasjournal} \bibliography{ms}

\pagebreak
\begin{deluxetable}{ll} 
\tabletypesize{\scriptsize}
\tablecaption{List of parameters for the energy budget investigation} 
\tablehead {\colhead{Parameter} & \colhead{Description} } 

\startdata
TE &  Thermal energy  (erg) \\
KE &   Kinetic energy  (erg) \\
L$_r$ &  Radiative loss energy (erg) \\
F$_c$ &  Thermal conduction energy (erg) \\
H &  Heating energy  (erg) \\
$\tau_{rad}$ & Radiative loss timescale (sec$^{-1}$) \\
$\tau_{cond}$ & Thermal conduction timescale (sec$^{-1}$) \\
L$_{rad}$ & Radiative loss rate (erg cm$^{-3}$sec$^{-1}$) \\
L$_{a}$ &Cooling rate by adiabatic expansion (erg cm$^{-3}$sec$^{-1}$) \\
dq & Cooling rate by thermal conduction (erg cm$^{-3}$sec$^{-1}$) \\
H$_r$ & Heating rate (erg cm$^{-3}$sec$^{-1}$)  
\enddata 
\label{tb:list}
\end{deluxetable}

\begin{deluxetable}{cccccccc} 
\tabletypesize{\scriptsize}
\tablecaption{Geomerical parameters for the estimation of the energy budget} 
\tablehead{
\colhead{Eruption} & \colhead{Time} & \colhead{{Depth}} & \colhead{Length} & \colhead{{Height}}  & \colhead{Velocity} &  \colhead{T$_0$} & \colhead{$\Delta$t}   \\
     & UT & \multicolumn{3}{c}{km} & km/s & MK & sec
}
\startdata
Prominence$^a$ & 17:36    & 5234   &    68994  &                  &         &            &            \\
Prominence$^b$ & 17:47    & 7918   &  133907  &     49750   &  40   & 0.08    &   648    \\
Loop                   & 17:47    & 25005  &  126768  &     45131   &  33   &    6      &   648    \\
Loop                   & 17:58    & 23071  &  161533  &     61838   &  33   &   10     &   684    \\
Loop                  & 18:08    & 23868  &  234976  &     95337    &  85  &    8     &   600    \\
\enddata 
\label{tb:parameters}
\tablecomments{\\ 
$^a$ Prominence as seen absorption features \\
$^b$ Prominence as seen emission features \\
 }
\end{deluxetable}

\begin{deluxetable}{ccccccccccccccccc} 
\tabletypesize{\scriptsize}
\tablecaption{Energy budget of the erupting prominence and loops} 
\tablehead{
\colhead{Eruption} & \colhead{Time} & \colhead{T} & \colhead{TR$^c$} & \colhead{n} & \colhead{Mass} &
\colhead{TE} &  \colhead{KE} & \colhead{L$_r$} & \colhead{F$_c$} & \colhead{H}  & \colhead{$\tau_{rad}$} & \colhead{$\tau_{cond}$}   &
\colhead{L$_{rad}$}  & \colhead{L$_{a}$} & \colhead{dq} & \colhead{H$_r$} \\
& (UT) & \multicolumn{2}{c}{(MK)} & (10$^{9}$cm$^{-3}$) & (10$^{14}$g) & \multicolumn{5}{c}{(10$^{28}$erg)} & \multicolumn{2}{c}{(10$^3$sec)} & \multicolumn{4}{c}{(10$^{-2}$erg cm $^{-3}$sec$^{-1}$)} 
}
\startdata
Prom$^a$ & 17:36    & 0.08  &   & 7.6-176  & 0.2-5.2 &                &              &              &           &                 &           &             &              &                   &         &        \\
Prom$^b$ & 17:47    & 2.5    &  0.5-3 & 13-25  & 1.7-3.2 &   17-30    & 0.1-0.3  &  6.9-23   & 0.8  &  19-44      &   0.5-1     & 65-120    & 1.3-4.4  & 0.1  & 0.0 & 5-10 \\
Loop                   & 17:47    & 10      &  6-14 & 9.0-16 & 11-20   & 416-737  & 0.6-1.0  &  18-56   &  111  &  357-566   & 5.3-9.5 & 1.2-2.1  & 0.4-1.1  & 3.1-5.6  & 2.8   & 9-14 \\
Loop                   & 17:58    & 8        &  4-11 & 8.6-16  & 11-21   & 346-631  & 0.6-1.1  &  23-78   &   36   &  129-233   & 3.5-6.3 & 3.3-6.0 & 0.4-1.3 & 3.2-5.8  & 0.8   & 3-5 \\
Loop                   & 18:09    & 4        &  1-14 & 6.5-12 &  14-24  & 205-362  & 4.9-8.7  &  15-46  &    2.0   &  46-96 & 2.9-5.2 &  30-53   & 0.2-0.6 & 3.2-5.7  & 0.0   & 0.7-1.5 \\
\enddata 
\label{tb:energy1}
\tablecomments{\\
$^a$ Prominence as seen absorption features \\
$^b$ Prominence as seen emission features \\
$^c$ Temperature range for the estimation of the densities  \\
 }
\end{deluxetable}

\begin{deluxetable}{cccccccccccccccc} 
\tabletypesize{\scriptsize}
\tablecaption{Energy budget of the erupting loops  (6-14 MK) } 
\tablehead{
\colhead{Eruption} & \colhead{Time} & \colhead{T} & \colhead{TR} & \colhead{n} & \colhead{Mass} &
\colhead{TE} & \colhead{L$_r$} & \colhead{F$_c$} & \colhead{H}  & \colhead{$\tau_{rad}$} & \colhead{$\tau_{cond}$}   &
\colhead{L$_{rad}$}  & \colhead{L$_{a}$} & \colhead{dq} & \colhead{H$_r$} \\
& (UT) & \multicolumn{2}{c}{(MK)} & (10$^{9}$cm$^{-3}$) & (10$^{14}$g) & \multicolumn{4}{c}{(10$^{28}$erg)} & \multicolumn{2}{c}{(10$^3$sec)} & \multicolumn{4}{c}{(10$^{-2}$erg cm$^{-3}$sec$^{-1}$)} 
}
\startdata
Loop                   & 17:58    & 10    & 6-14      &   8.5-15  & 11-21   & 430-775  &   19-60   &   78   &  239-389   & 5.5-9.9 & 1.9-3.4 & 0.3-1.0 & 3.1-5.7  & 1.7   & 5.2-8.4 \\
Loop                   & 18:09    & 10    & 6-14      &   4.3-7.6 &  9.0-16  & 340-592  &   6.5-20  &    51   &  224-360 & 11-20 &  2-3.5   & 0.1-0.3 & 2.7-4.6  & 0.8   & 3.6-5.7 \\
\enddata 
\label{tb:energy2}
\end{deluxetable}

\pagebreak

\begin{figure}
\epsscale{1.0}\plotone{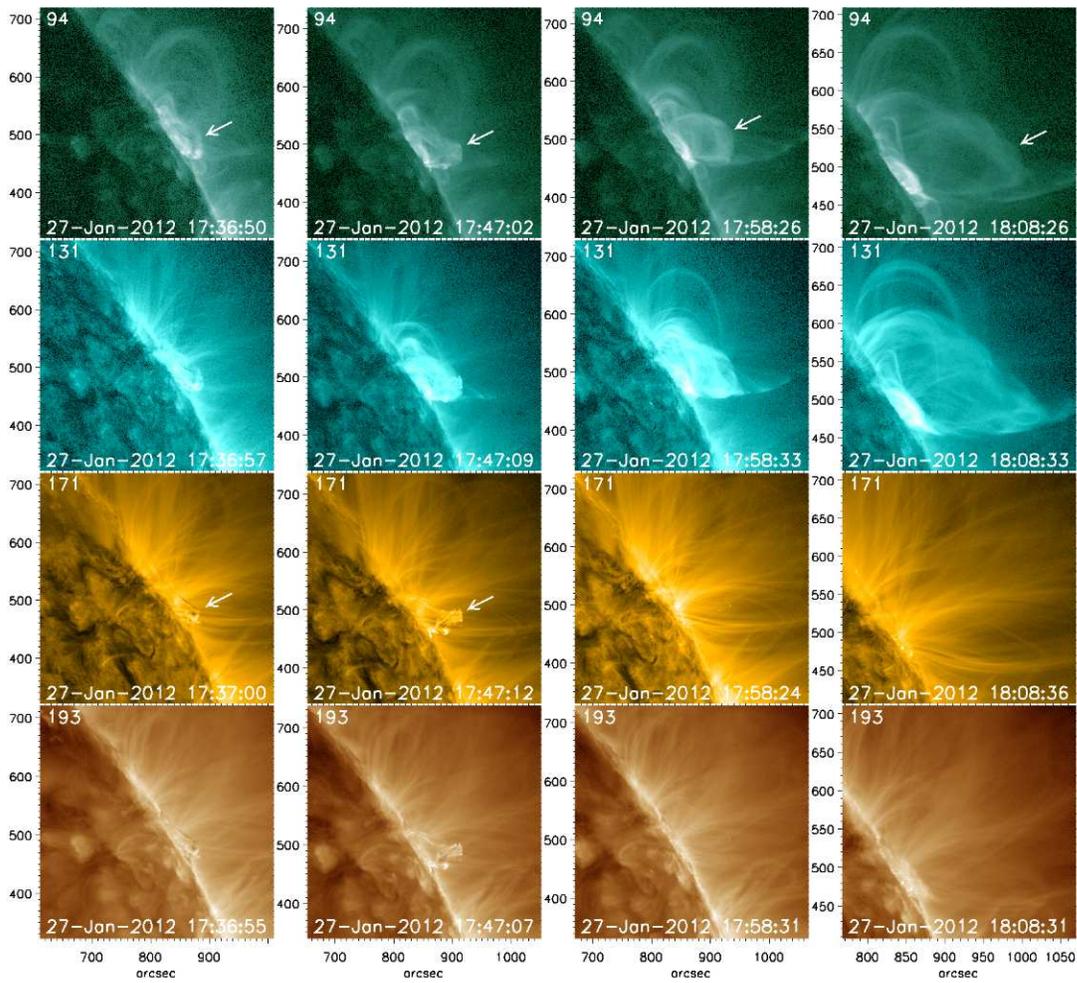}
\caption{Eruptions of the prominence and loops on 2012 January 2012 observed by AIA and XRT. The XRT observations are shown only at three times. 
Arrows in the AIA/94 \AA\ and XRT/Be$\_$thin observations represent the erupting loops. 
Arrows in the first and second columns in 171 \AA\ observations represent the erupting prominences. 
The prominence with a writhe structure of the prominence is seen in the second column. } 

\label{fig:obs}
\end{figure}
\addtocounter{figure}{-1}
\begin{figure}
\epsscale{1.0}\plotone{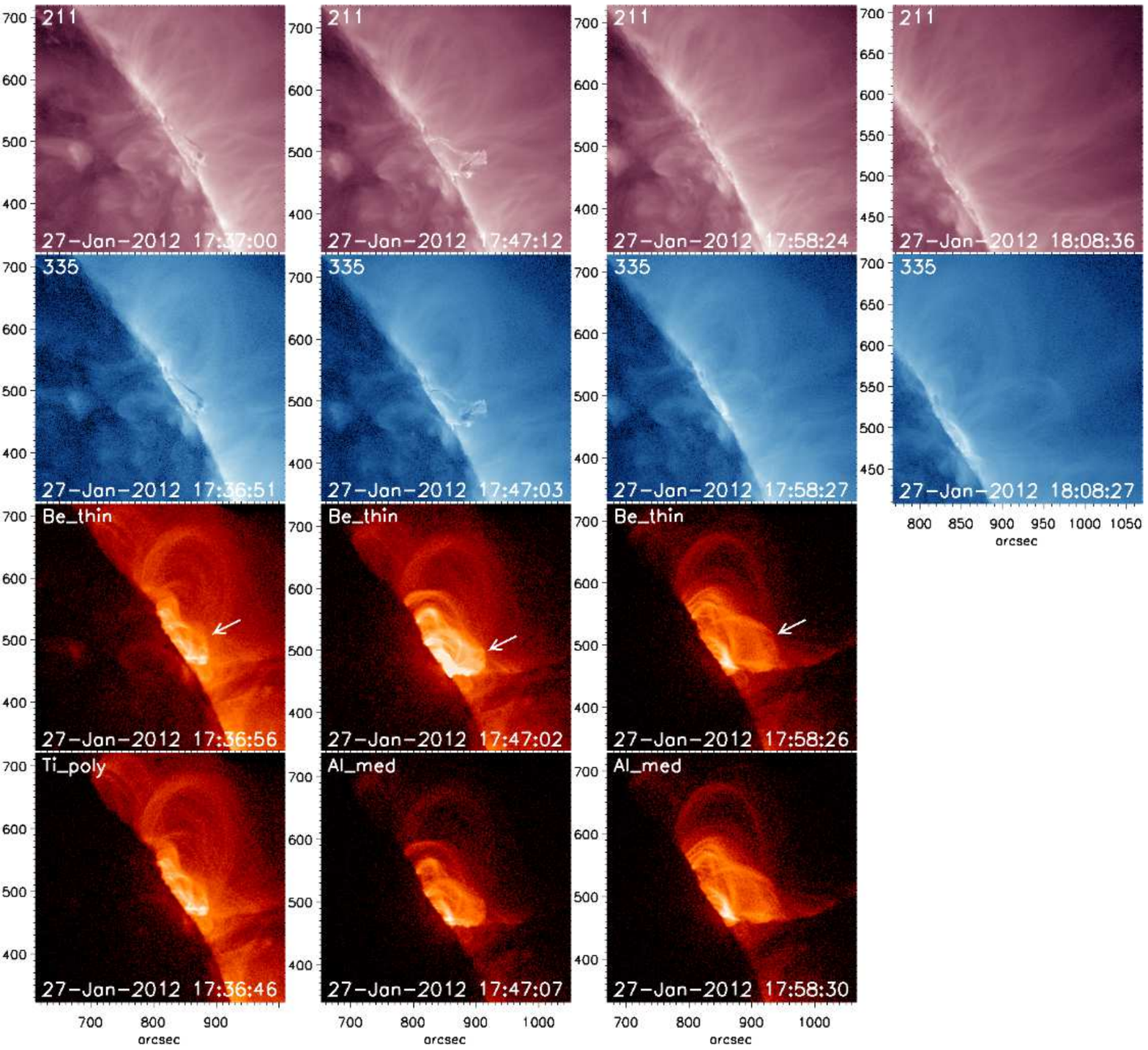}
\caption{(continued)} 
\label{fig:obs}
\end{figure}

\begin{figure}
\centering
\includegraphics[width=100mm]{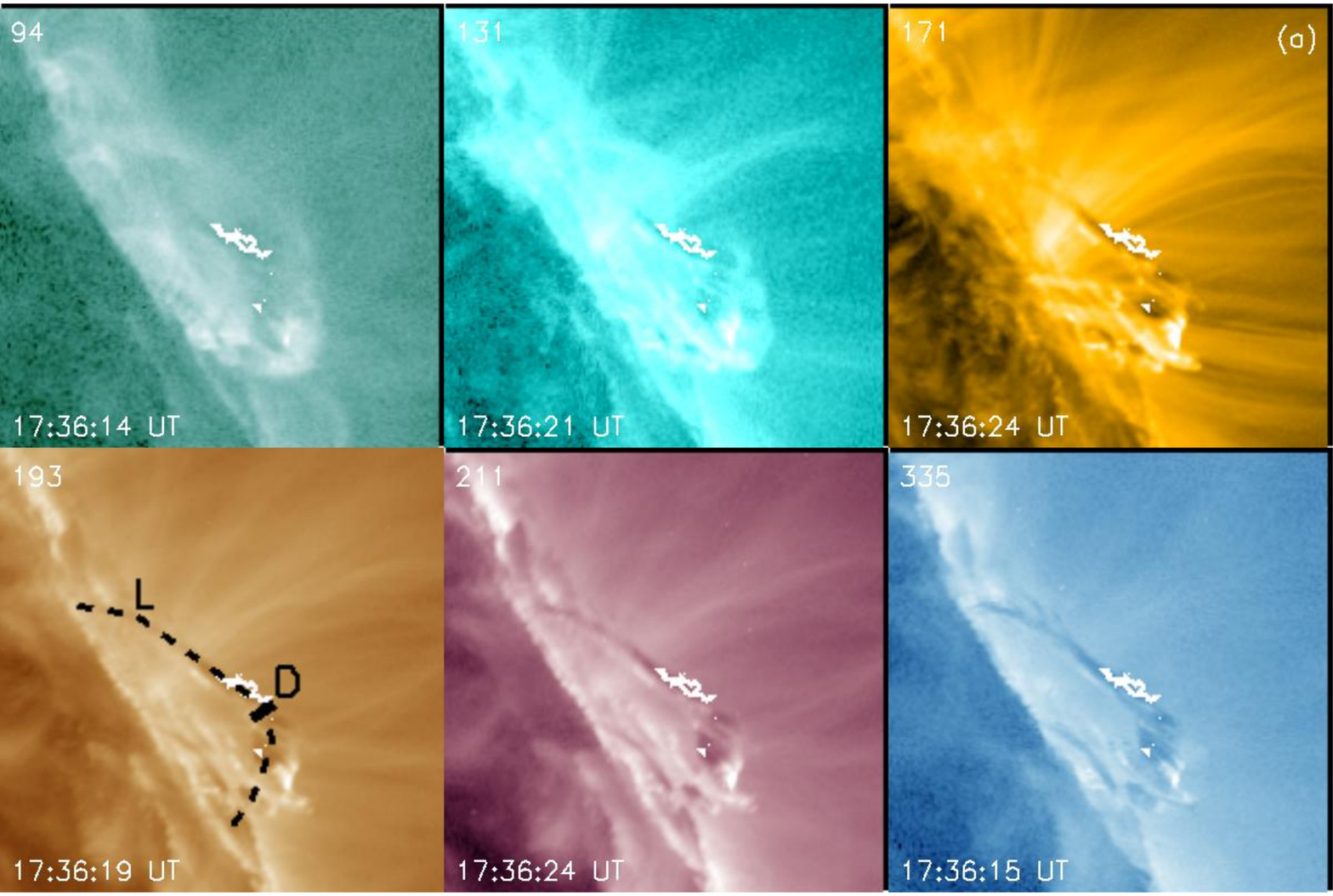}
\includegraphics[width=100mm]{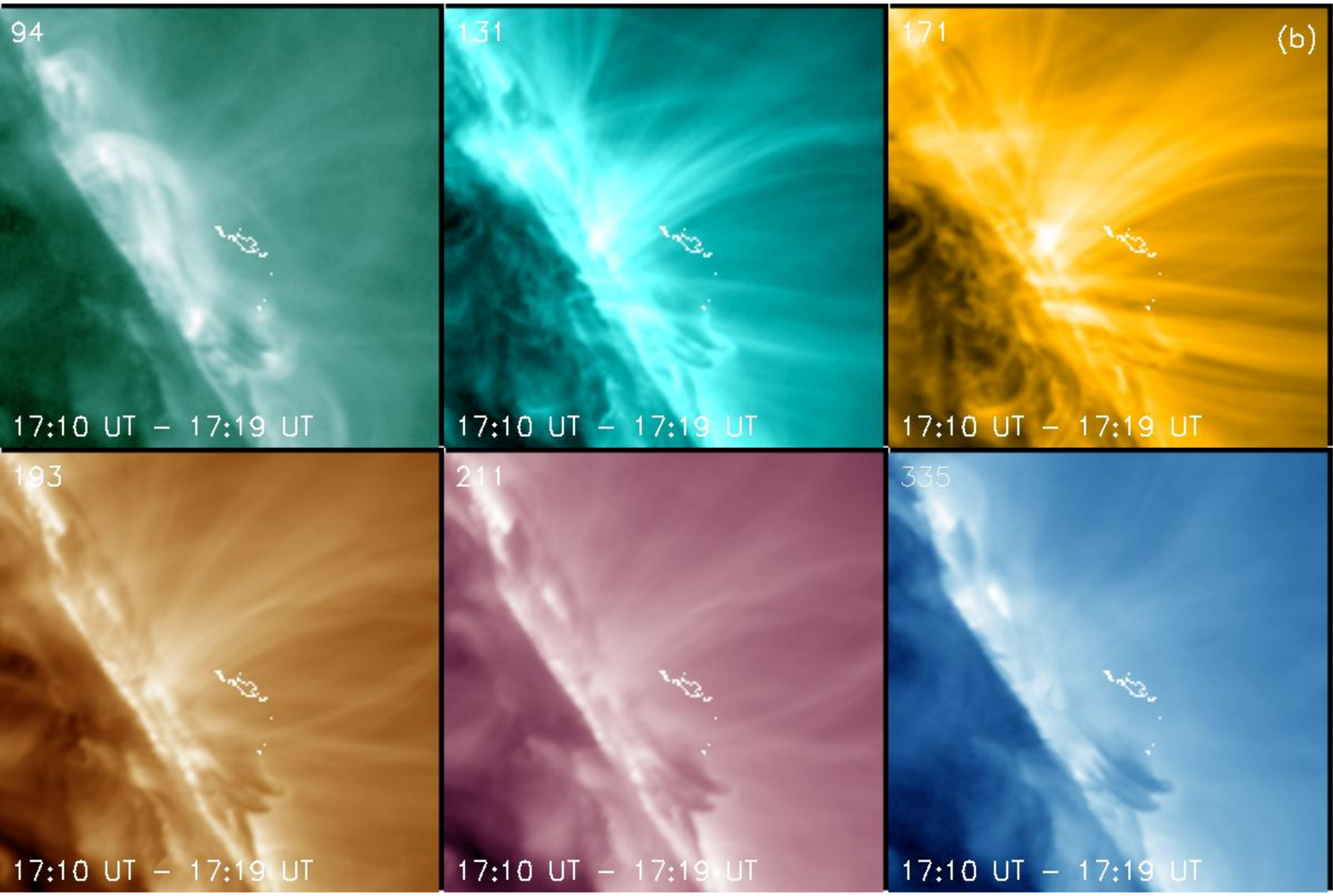}
\caption{(a) AIA images with selected absorption features shown as contours. (b) Averaged emission between 17:10 UT and 17:20 UT with the contours in (a). }
\label{fig:ratios}
\end{figure}

\begin{figure}
\epsscale{1.0}\plotone{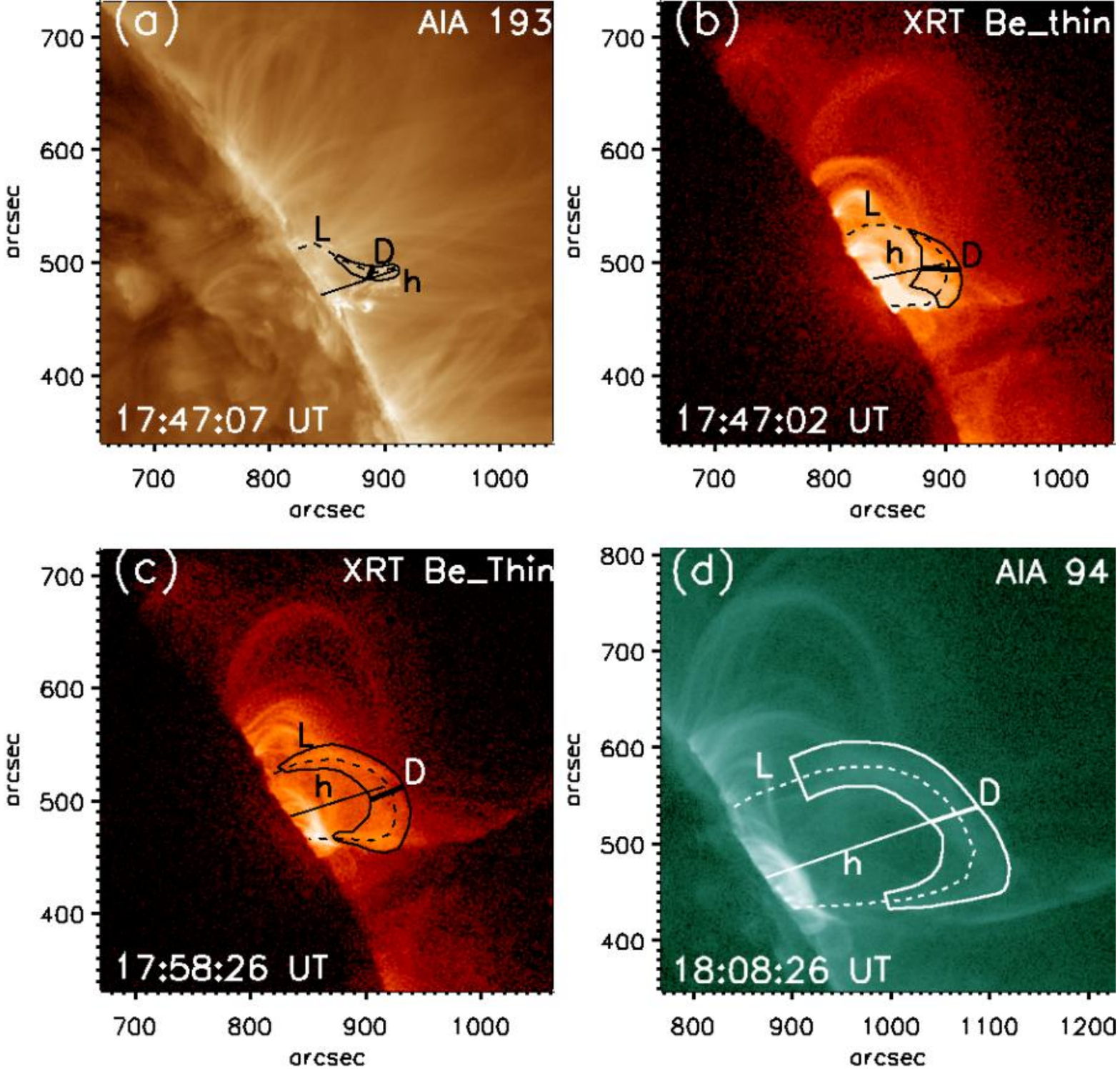}
\caption{Selected regions and assumed geometries of the erupting prominence and loops. Erupting prominence (a) and loop (b) at 17:47 UT, Erupting loop at 17:58~UT (c) and 18:08~UT (d). The L, D, and H represent the assumed lengths, widths (line of sight depths), and heights, respectively.} 
\label{fig:regions}
\end{figure}

\begin{figure}
\centering
\includegraphics[width=100mm]{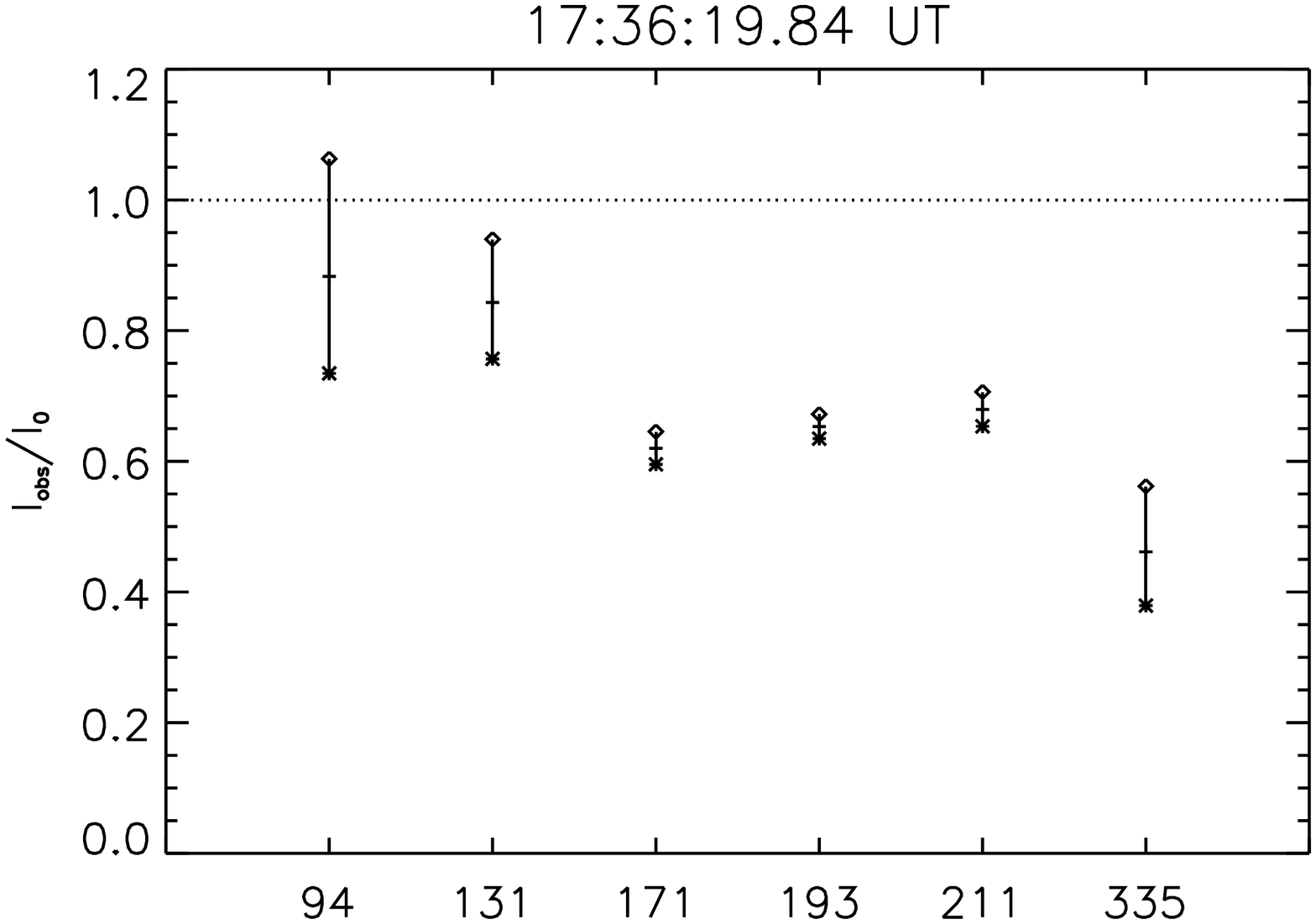}
\caption{ $I_{obs}/I_{0}$ of the regions enclosed by contours.}
\label{fig:ratios}
\end{figure}

\begin{figure}
\centering
\includegraphics[width=100mm]{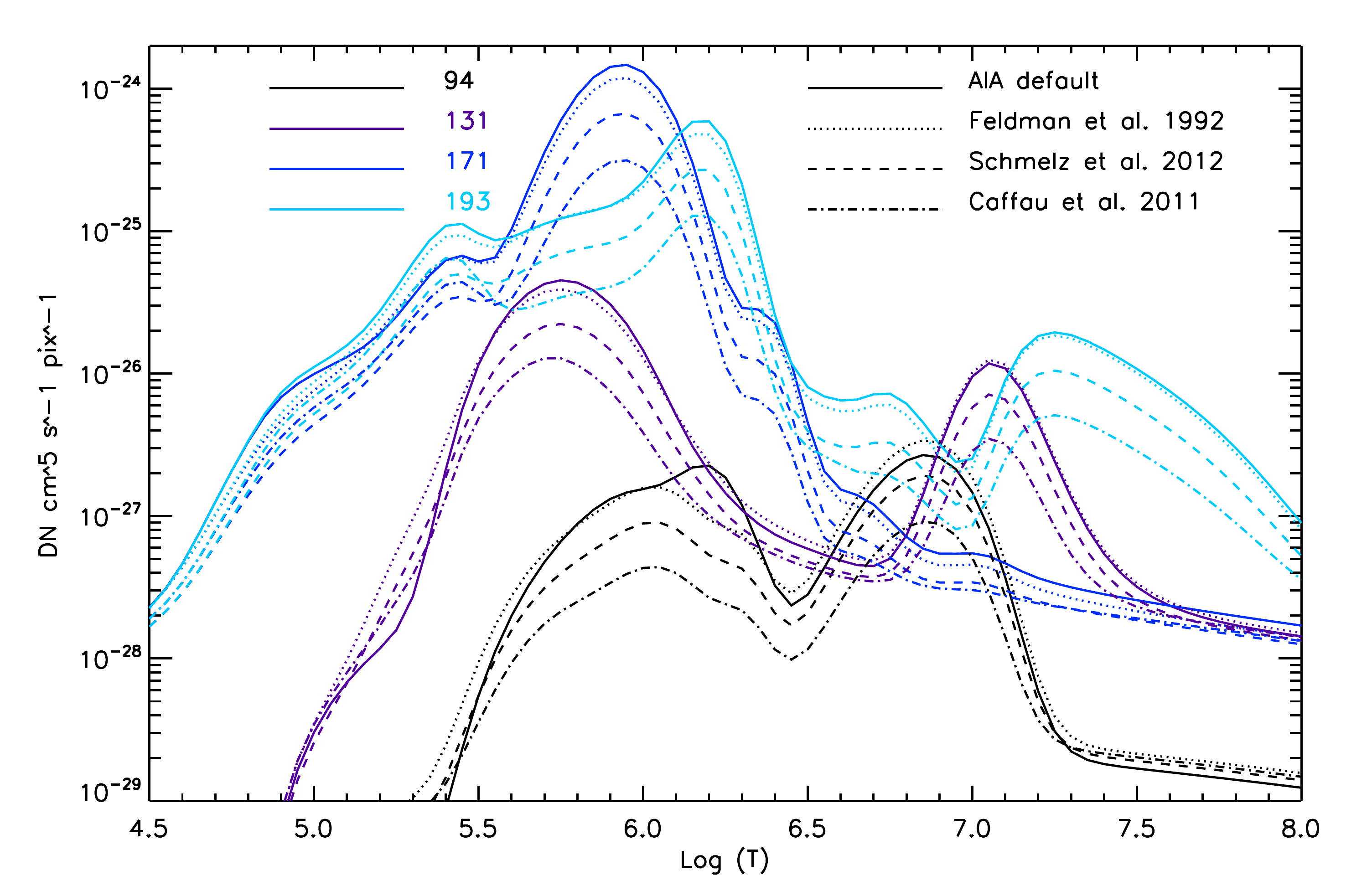}
\includegraphics[width=100mm]{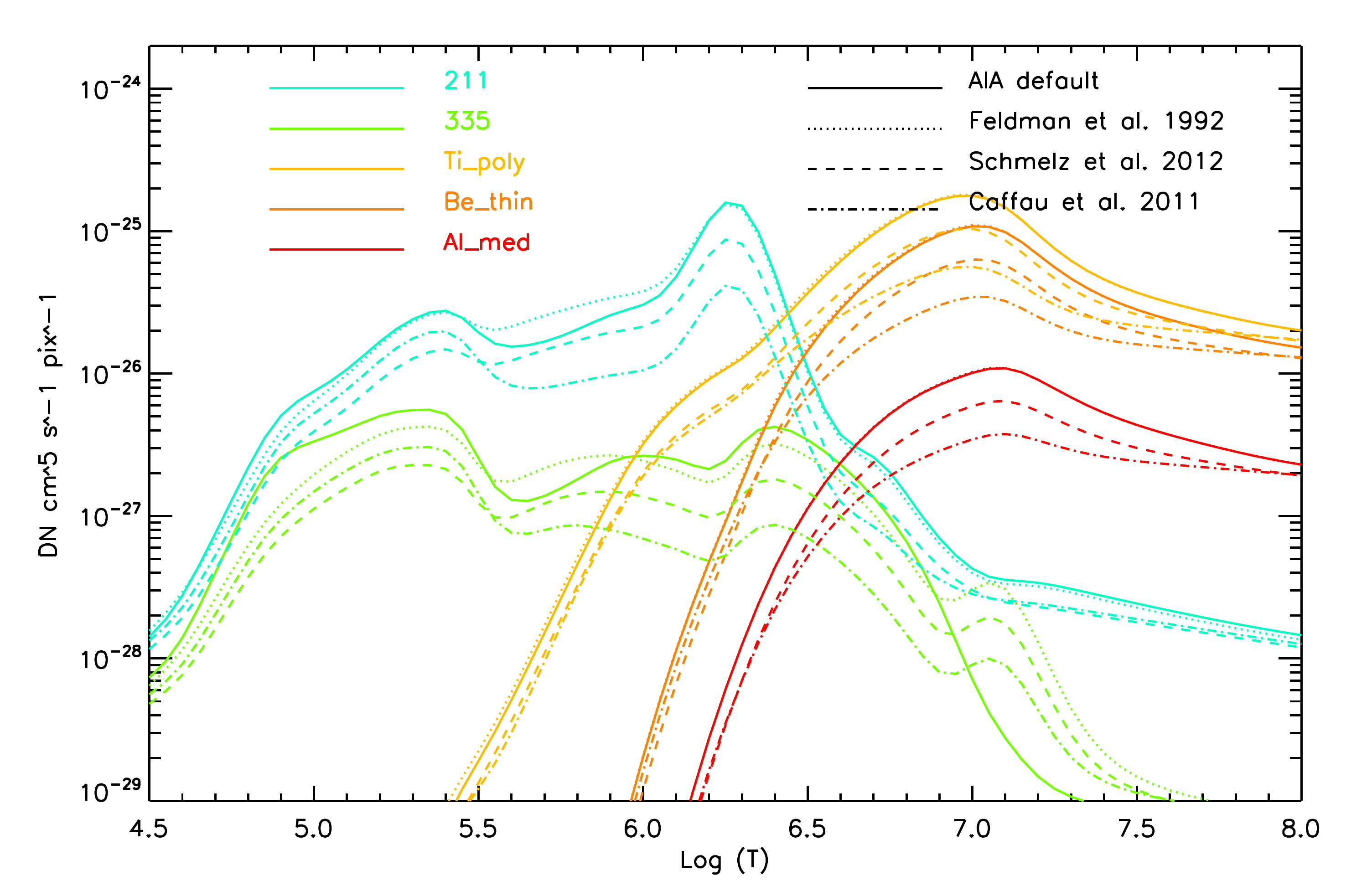}
\caption{Temperature response functions made with various synthetic spectra for the SDO/AIA and Hinode/XRT} 
\label{fig:response}
\end{figure}

\begin{figure}
\epsscale{1.0}\plotone{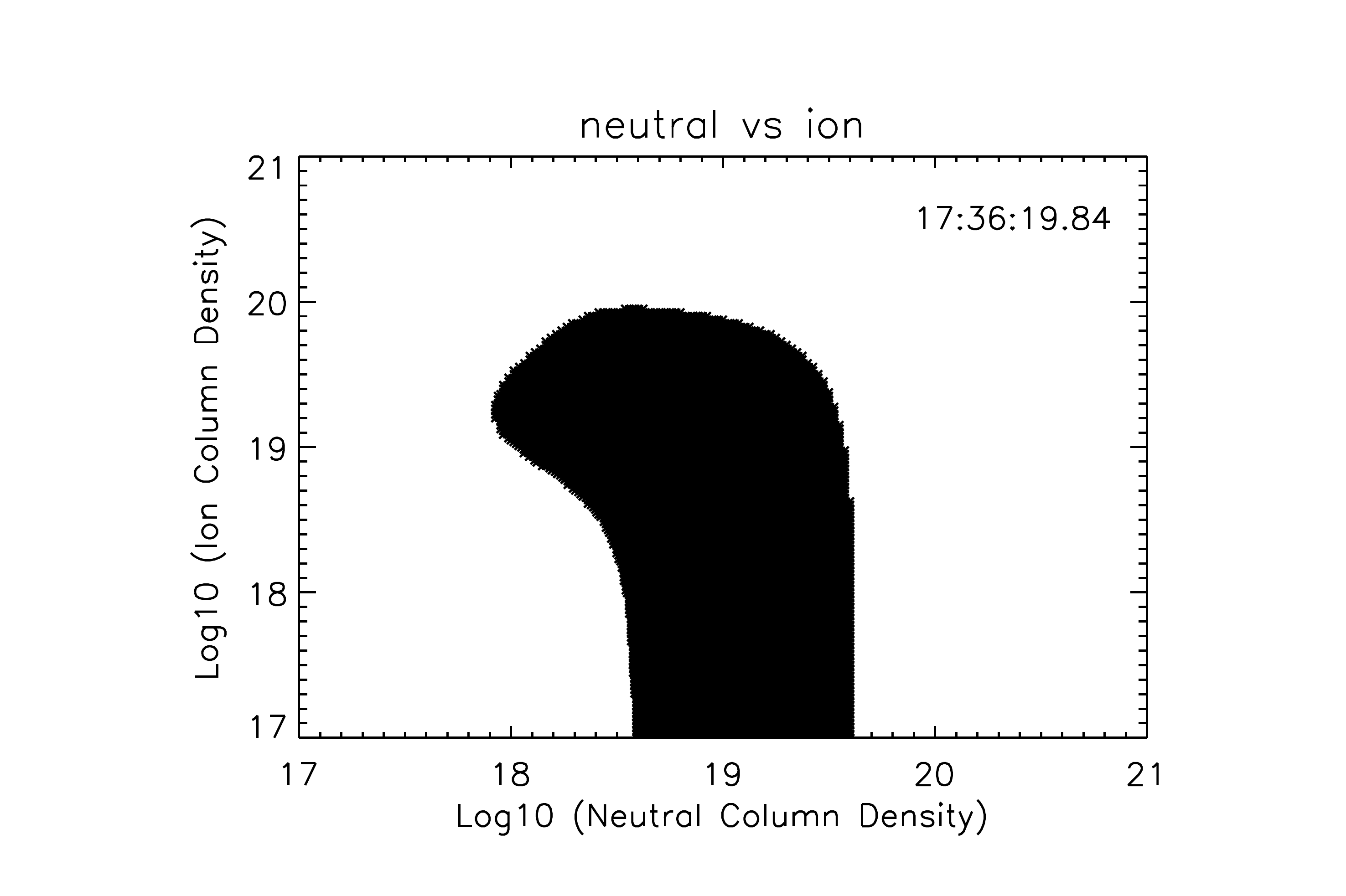}
\caption{Constraint of the column density of the selected region from the absorption features in Figure~2.} 
\label{fig:regions}
\end{figure}

\begin{figure}
\centering
\includegraphics[width=150mm]{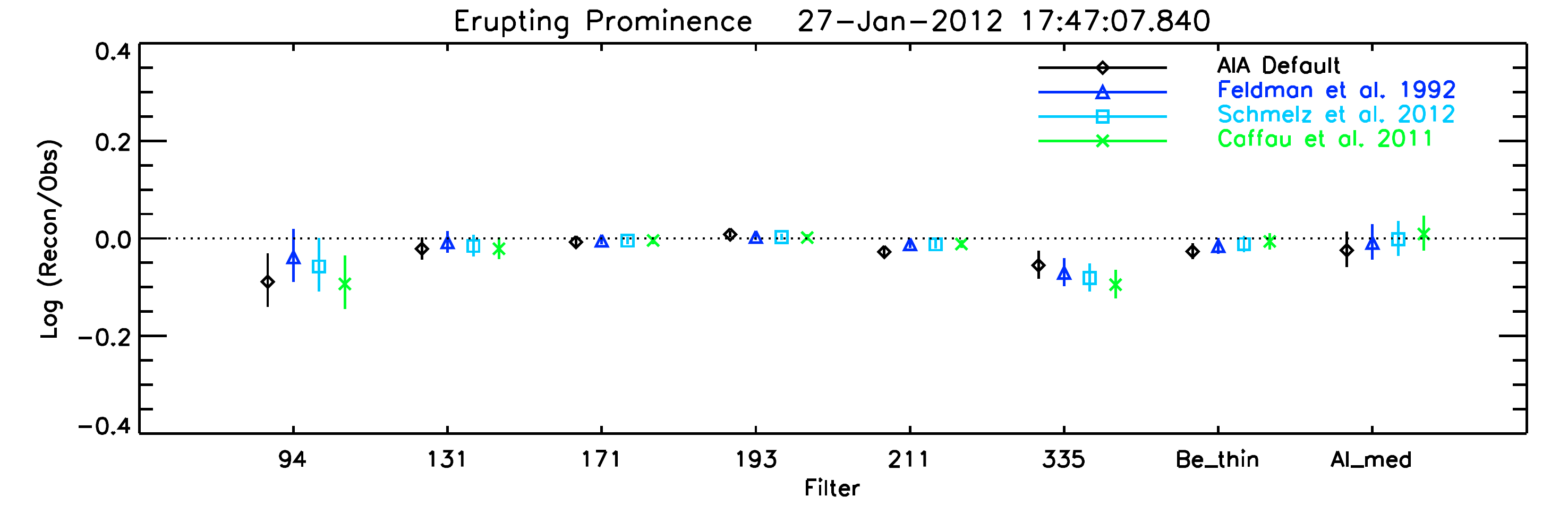}
\includegraphics[width=150mm]{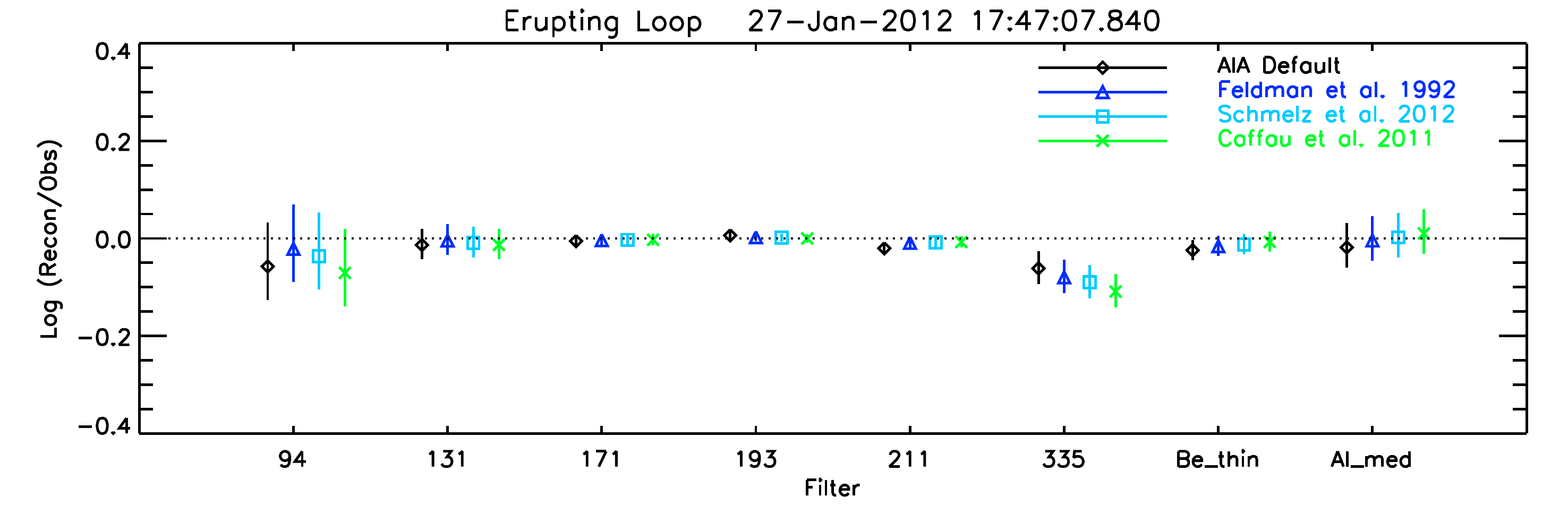}
\includegraphics[width=150mm]{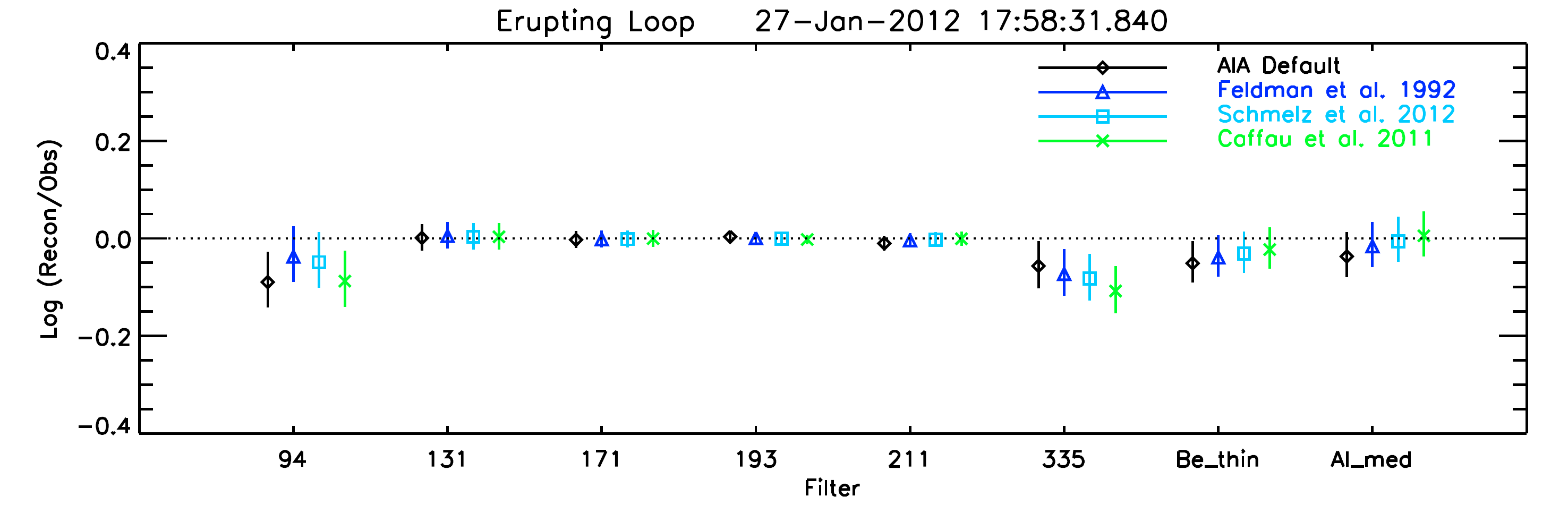}
\includegraphics[width=150mm]{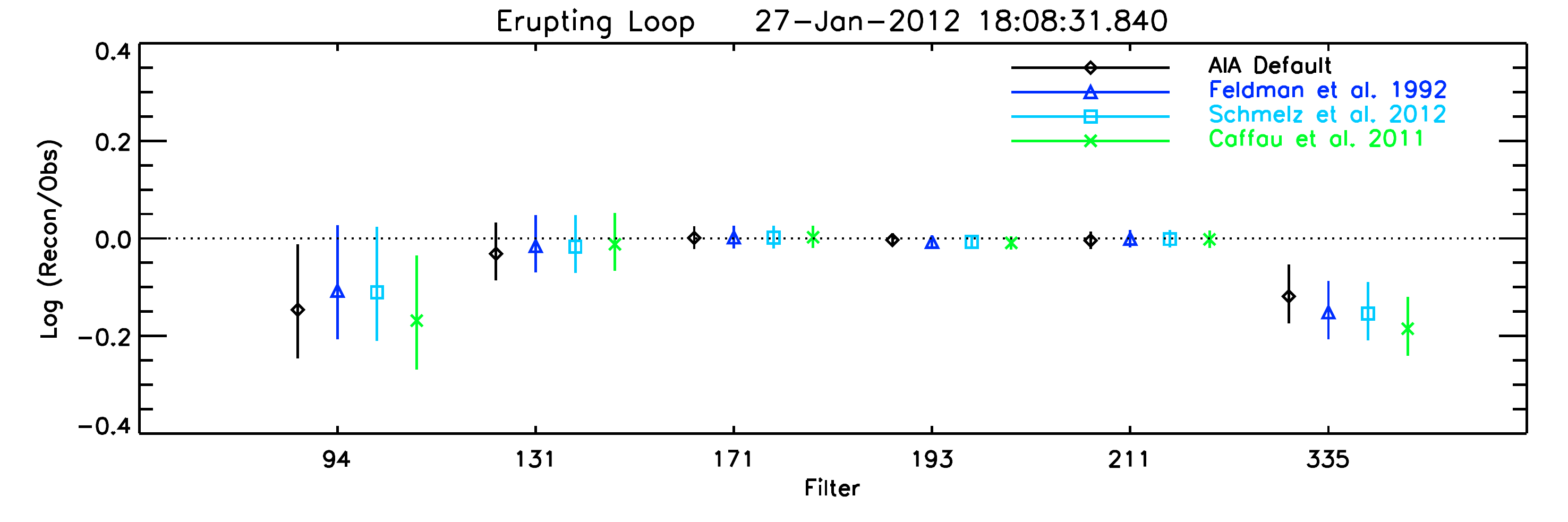}
\caption{DEM reconstruction errors for the erupting prominence and loops.} 
\label{fig:error}
\end{figure}

\begin{figure}
\centering
\includegraphics[width=150mm]{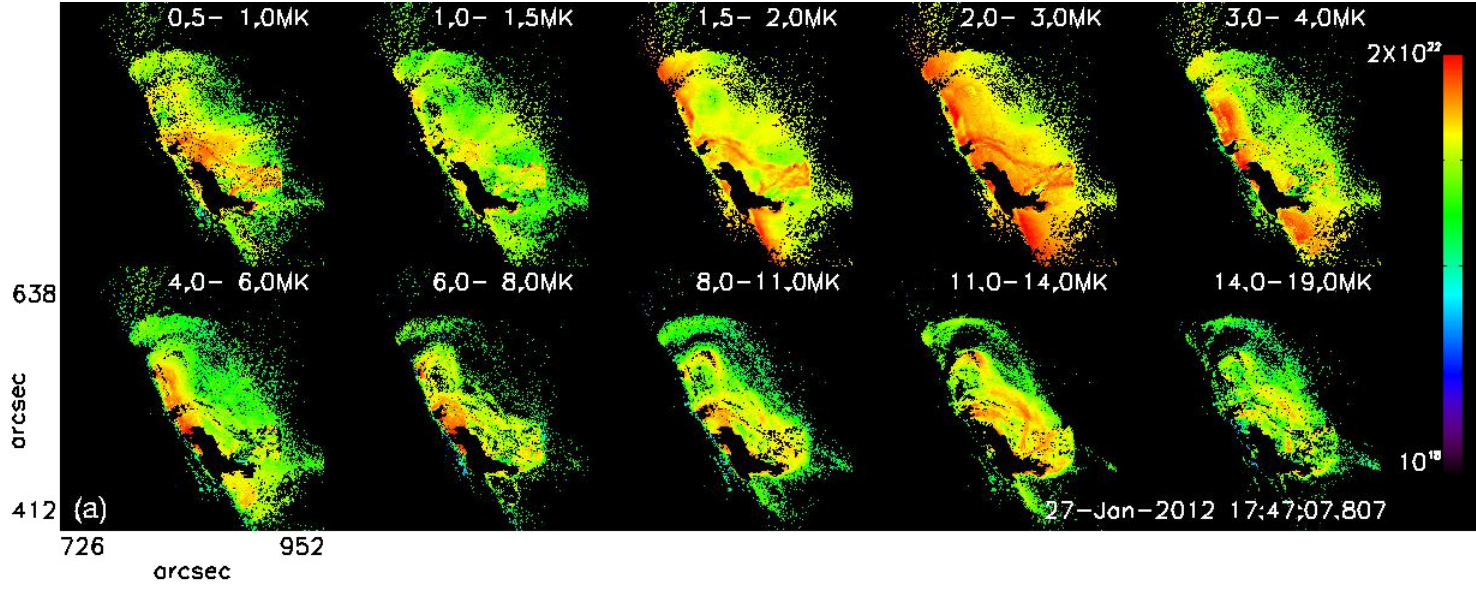}
\includegraphics[width=150mm]{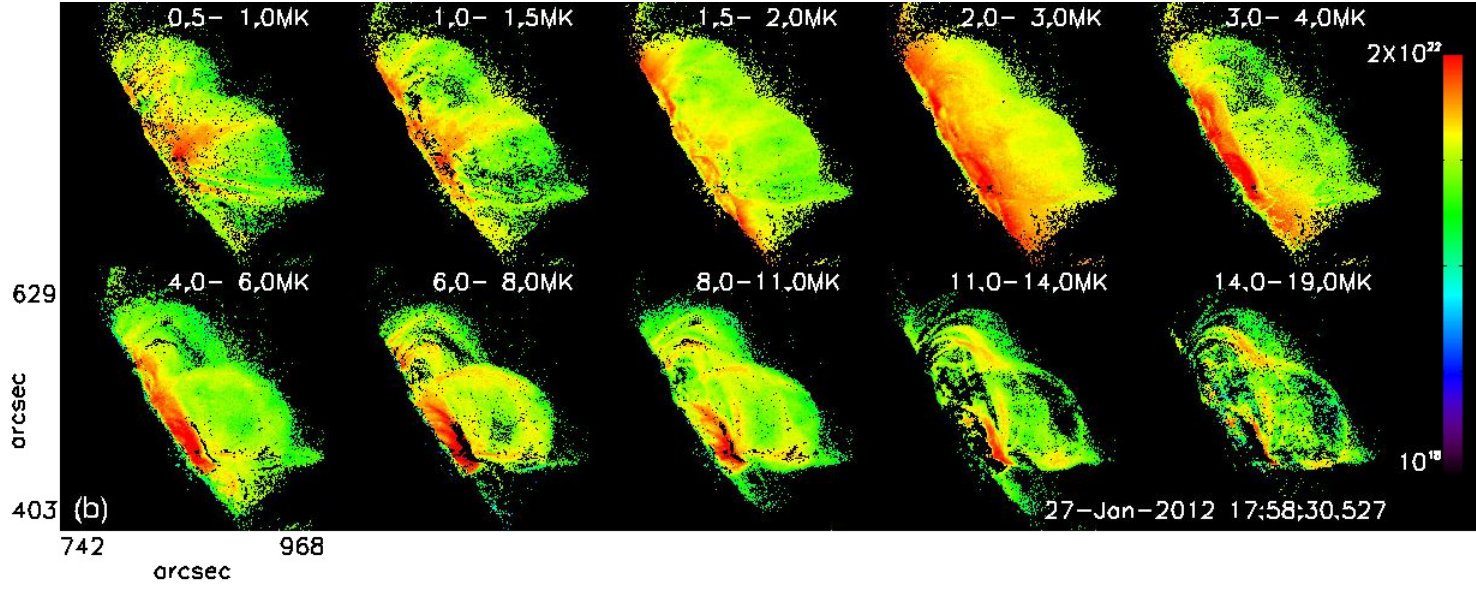}
\includegraphics[width=150mm]{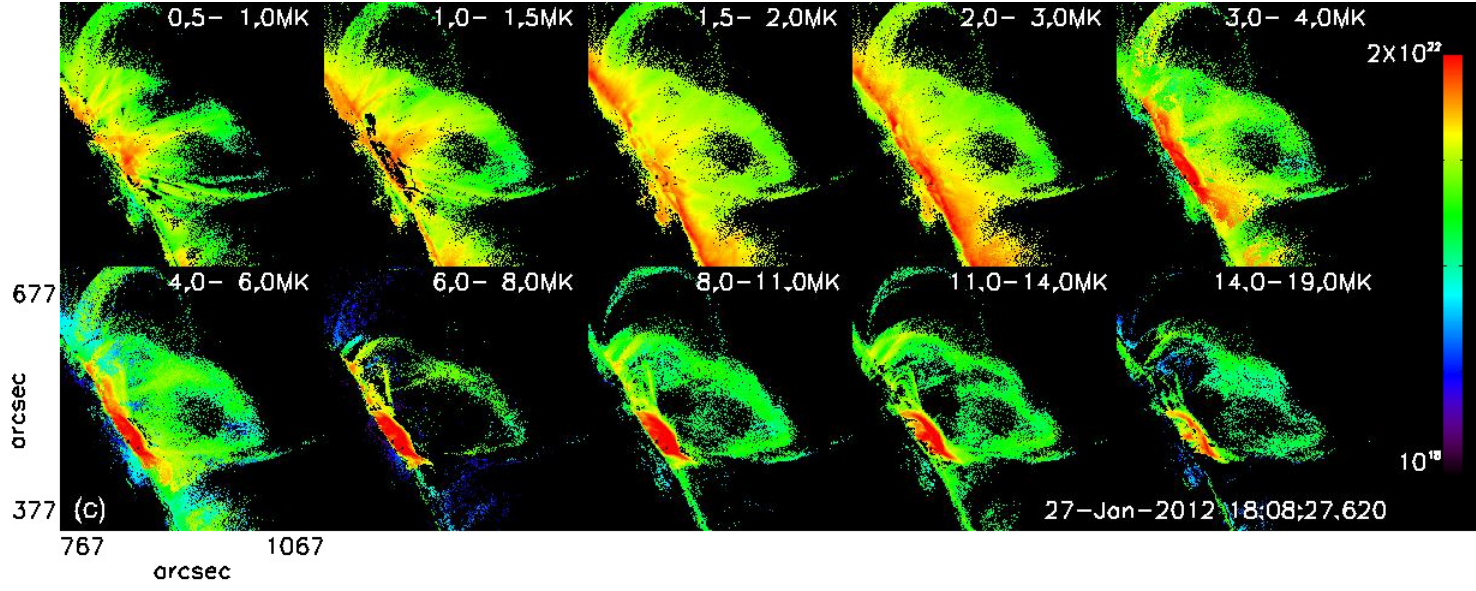}
\caption{DEM maps of the erupting prominence and loops at 17:47 UT (a), 17:58 UT (b), and 18:08 UT (c).} 
\label{fig:error}
\end{figure}

\begin{figure}
\centering
\includegraphics[width=90mm]{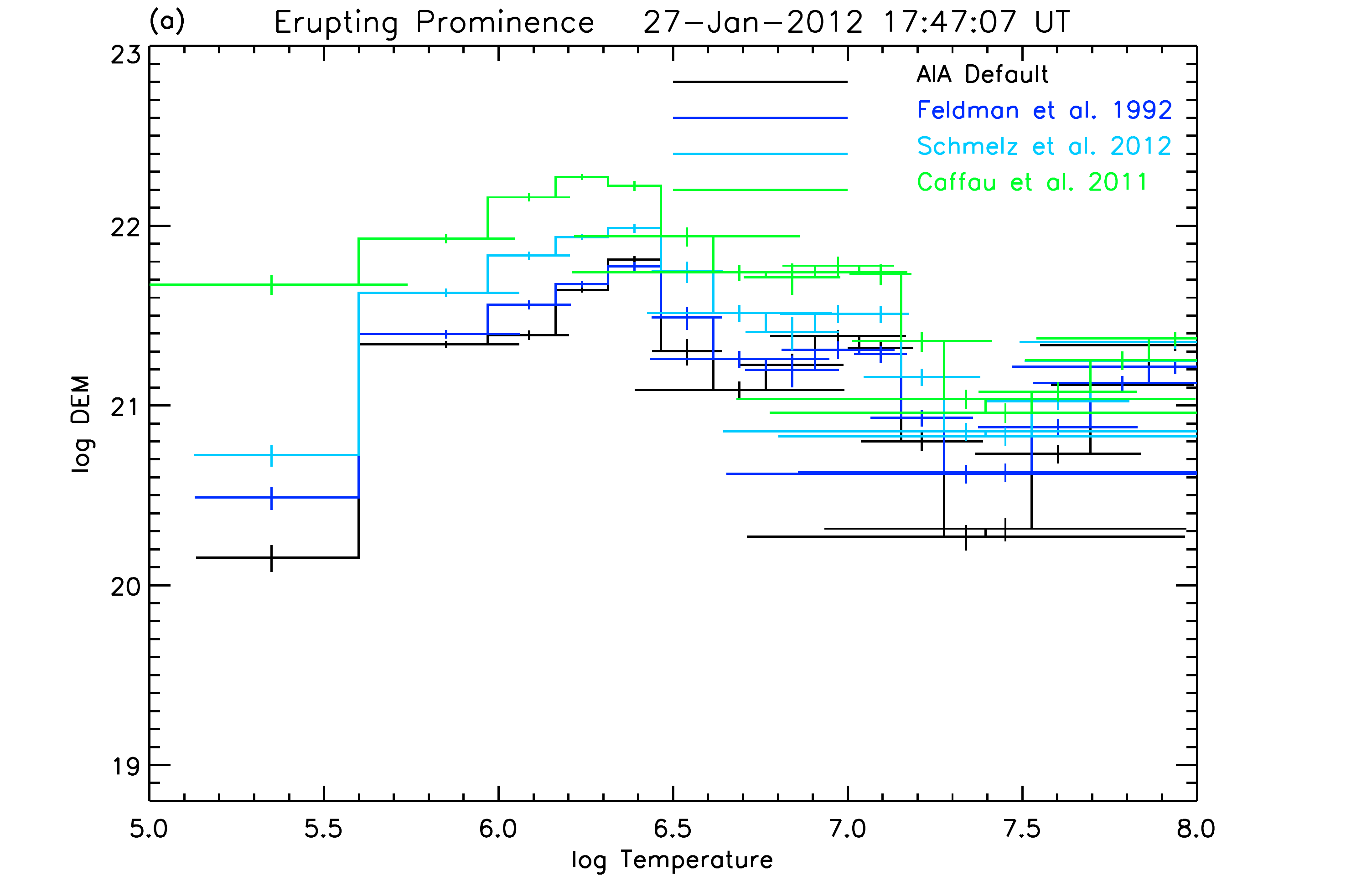} 
\includegraphics[width=90mm]{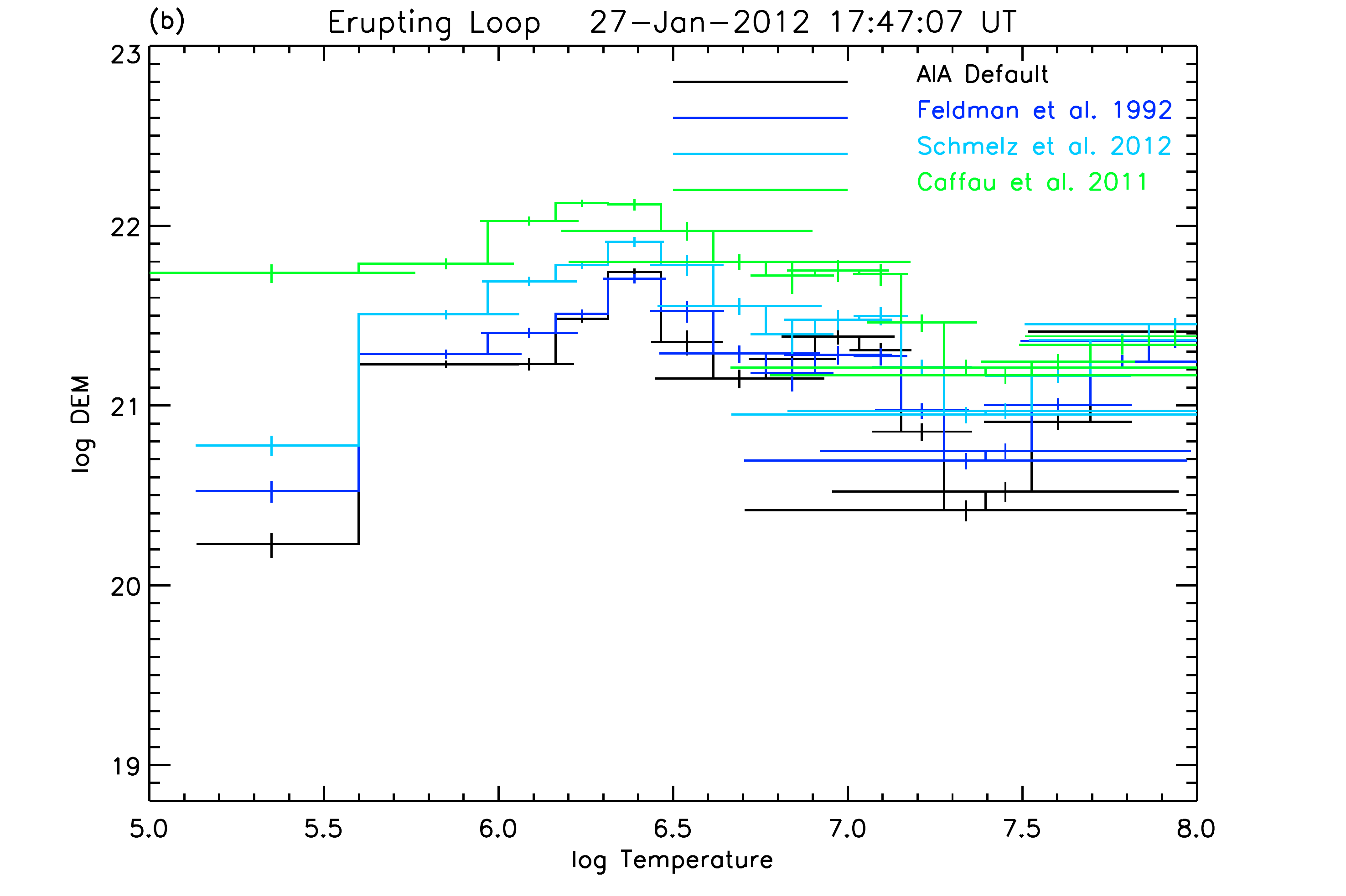} 
\includegraphics[width=90mm]{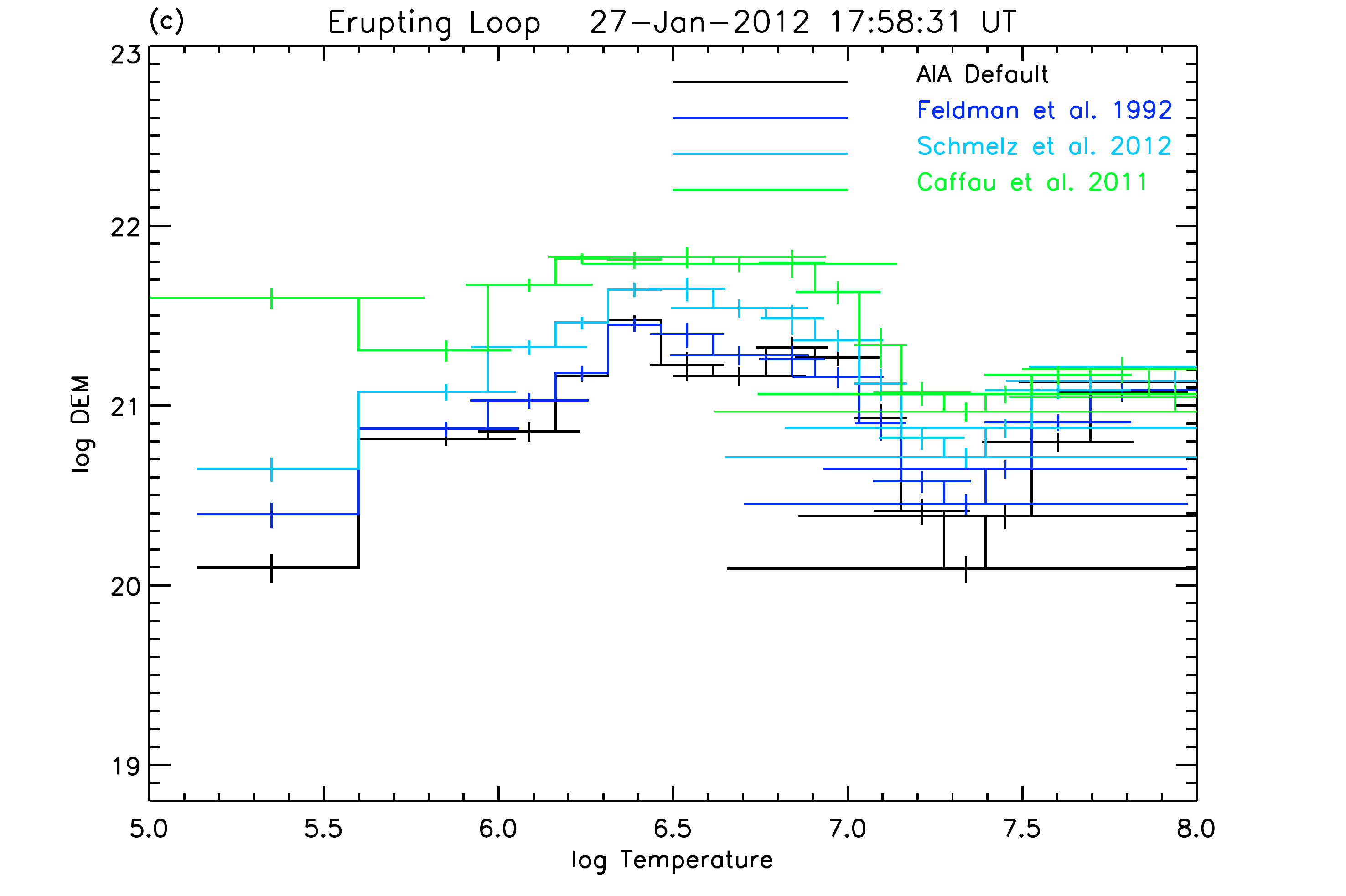} 
\includegraphics[width=90mm]{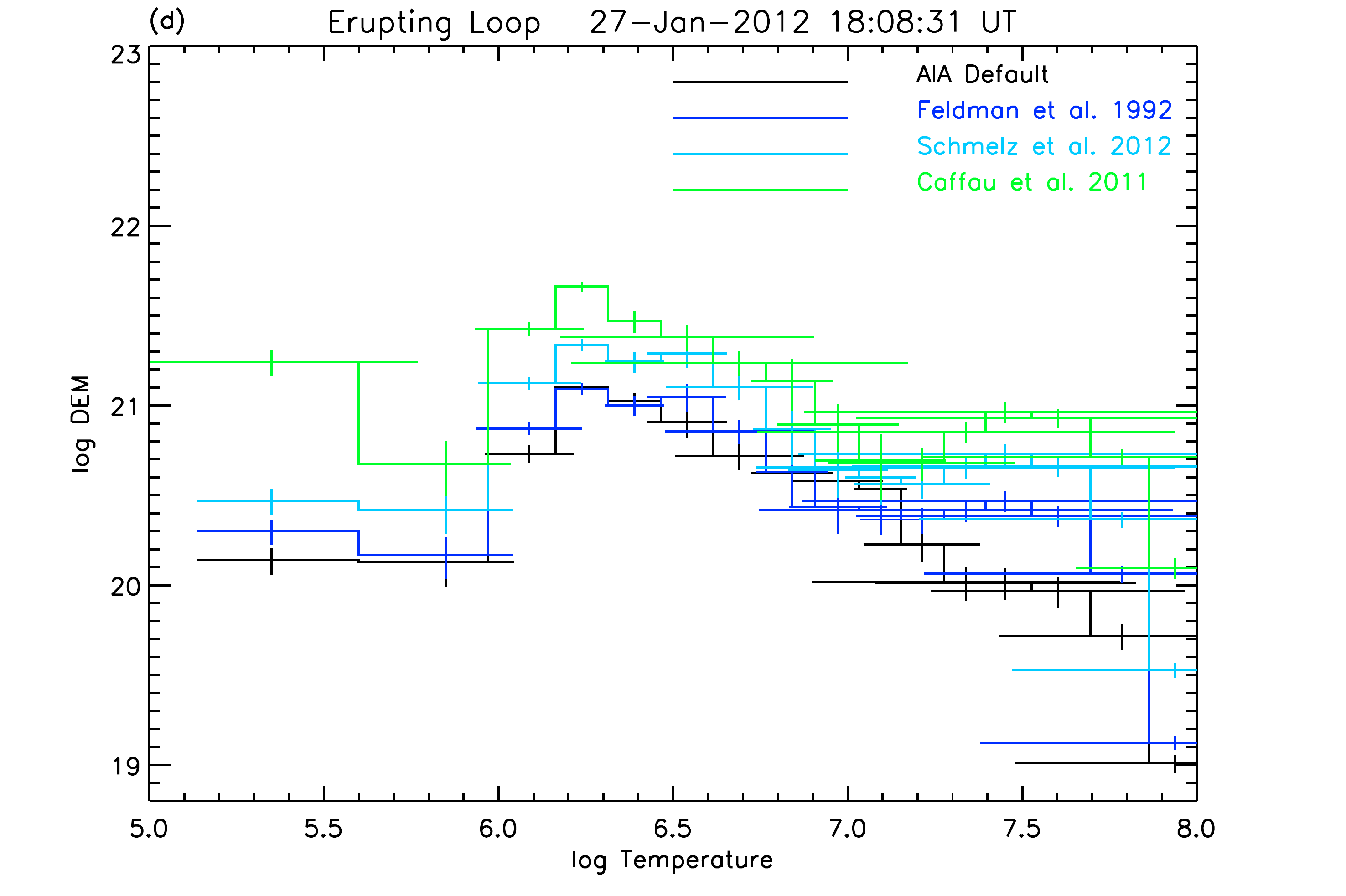} 
\caption{DEMs for the erupting prominence and loops by using various abundances.} 
\label{fig:dem}
\end{figure}

\end{document}